\documentclass{ws-ijmpa}

\usepackage[super]{cite}
\usepackage{xcolor}
\usepackage[verbose,hypertexnames=false]{hyperref}
\usepackage{soul}
\hypersetup{colorlinks=false,allbordercolors=blue,pdfborderstyle={/S/U/W 1}}

\newcommand{\chp}[1]{\mbox{$\stackrel{\wedge}{#1}$}}

\newcommand{\vct}[1]{\mbox{$\stackrel{\rightarrow}{#1}$}}

\newcommand{\slas}[1]{\mbox{${{#1} \!\!\! /}$}}

\begin{document}

\markboth{John Mashford}{Spectral QCD running coupling}

%
\catchline{}{}{}{}{}
%

\title{Computation of the one-loop spectral QCD running coupling using covariant spectral regularization}

\author{John Mashford}

\address{School of Mathematics and Statistics, University of Melbourne,\\
Victoria, 3010, Australia\\
mashford@unimelb.edu.au}

\maketitle

\begin{history}
\received{Day Month Year}
\revised{Day Month Year}
\accepted{Day Month Year}
\published{Day Month Year}
\end{history}

\begin{abstract}
The running strong coupling is an important quantity, indicating the strength of the strong force at any given energy. Methods described in the literature for its computation are essentially all defined with respect to the renormalization group equations and these equations are associated with the method of renormalization for dealing with infinities in Feynman integrals. The problem with the renormalization group equations is their prediction of the unphysical Landau pole which, for QCD occurs at an energy of the order of a few hundred MeV. The models described in the literature generally interlace high energy renormalization group predictions with modified low energy formulations. It would be desirable to have a method for the computation of the running strong coupling which is not {\em ad hoc} but is unified over the whole range of energies and is based on a single mathematically rigorous formulation which is guided by physical principles. In this paper we describe a  method using the regularization technique called covariant spectral regularization for which renormalization is not required.  We use covariant spectral regularization to compute the spectral QCD vacuum polarization tensor in order to determine and display the (one-loop) spectral QCD running coupling. The densities associated with the quark and gluon bubbles are computed without requiring renormalization and hence the spectral QCD vacuum polarization tensor is determined. We explain why we think these computations are sufficient to determine the QCD running coupling. It is found that the position space spectral QCD running coupling is an analytic function which does not manifest a Landau pole, but instead it manifests  what might be called a ``Landau peak",  and has the property known as ``freezing of $\alpha_s$" in the infrared. On enforcing the agreement of our results with two points in real $\overline{\text{MS}}$ data we determine a spectral bare strong coupling constant of $\alpha_b\approx(411)^{-1}$ which can be used in higher order QCD computations (this is to be compared with QCD using renormalization where the bare strong coupling is infinite). Thus it seems that one can conclude that, when analyzed using covariant spectral regularization, QCD is perturbative at all energies.
\end{abstract}


\ccode{PhySH subject headings: Perturbative QCD, Non-Abelian gauge theories}

\section{Introduction}

The strong force coupling, like the couplings of all the forces of nature, ``runs" with collision energy, that being the inverse of the distance to which a target is probed. The couplings of the gravitational, electromagnetic and weak forces are known to great accuracy but the coupling of the strong force (at, e.g., an energy $\tau=m_Z$ where $m_Z$ is the mass of the Z boson) is only known to an accuracy of a world average of about 0.8\% (see Ref. \citen{Snowmass21}). 

As is well known, the manner in which the strong force coupling ``constant" runs is different to, e.g., the manner in which the electromagnetic force coupling constant runs.

The running of the coupling in quantum chromodynamics (QCD) is an important phenomenon \cite{Achenbach,Dissertori,Hinchliffe,Prosperi} which needs to be taken into account in carrying out calculations in QCD \cite{Ducloue}. Since, using standard regularization techniques, the bare coupling is infinite, one must use the finite running coupling in QCD calculations.

In QCD the term strong coupling constant can refer to the quantity $g_s$ occurring in the expression defining the Feynman amplitude for any given process or else to the quantity $\alpha_s=\frac{g_s^2}{4\pi}$. The quantity being referred to is determined by which of the symbols $g_s$ or $\alpha_s$ is used. 

In QCD the running coupling is not an observable but, in the high energy domain where, due to asymptotic freedom \cite{Gross,Politzer} $\alpha_s\ll1$ so perturbative QCD (pQCD) is applicable, observables are calculated from $\alpha_s$  by expanding to $n^{\text{th}}$ order a series in $\alpha_s$. \cite{Wu_1} Thus $\alpha_s$ serves as an intermediate quantity between observables. In the low energy domain analytic expressions based on pQCD are, in various ways, supplemented with non-perturbative terms. \cite{Deur}

The running coupling constant $\alpha_s$ in QCD can be determined experimentally in a number of ways including deep inelastic scattering, $\tau$ decay and hadron-hadron scattering \cite{Hinchliffe,Snowmass21}. In addition it can be determined from lattice gauge theory computations \cite{Aoki}. In all cases its determination involves theoretical QCD calculations involving renormalization. The principal reason for this is that, unlike the case of, e.g., electrons in quantum electrodynamics (QED), free quarks have not been observed (except, possibly, in quark-gluon plasma) and the coupling must be inferred from the interactions of quarks bound in hadrons and calculable consequences, and the currently accepted calculation technique involves renormalization.

As a result of asymptotic freedom the behavior of $\alpha_s$ in the perturbative domain is known. However there are, in the literature, a number of different choices for $\alpha_s$ in the non-perturbative domain. ``There is no single, agreed prescription for defining an IR completion of QCD's running coupling". \cite{Deur2024} (Here, the term IR completion means a formulation of an expression for the running coupling which is valid (and non-divergent) at all energies including  low energies.)  ``The optimal choice of the definition of $\alpha_s(Q^2)$ at all scales is an unsettled question". \cite{Deur} 

The current standard method of computing the QCD running coupling in the high energy domain involves the use of the renormalization method in pQCD.
One of the main problems met with in pQCD, in fact in the use of renormalization in general, is the difficulty with setting the renormalization scale $\mu_r$, which arises through the ``dimensional transmutation" property of the renormalization method. Only after the renormalization scale has been set can predictions for physical observables be obtained. In the usual method a single $\mu_r$ is simply guessed at, and its value is then varied over an arbitrary range. \cite{Wu} This procedure is {\em ad hoc} and contradicts renormalization scheme invariance. Two approaches to getting around this problem are the principle of maximum conformality \cite{Wu} and the Brodsky-Lepage-Mackenzie  approach. \cite{Brodsky} 

The methods for setting the renormalization scale and defining the running coupling are all formulated in the context of the renormalization group involving the renormalization group equations. \cite{Gell-Mann,Callan,Symanzik} These equations predict a singularity, or divergence, in the running coupling called the Landau pole which, for QCD, is estimated to occur at an energy of the order of a few hundred MeV. Such a divergence is unphysical, it has not been observed either experimentally or in lattice simulations.

We have developed a method of regularization for quantum field theory (QFT) which we called spectral regularization and which is described in Refs. \citen{Springer,Symmetry,IJMPA,NPB,vertex} and briefly discussed in Section \ref{section:cov_spec_reg}. We now call this technique covariant spectral regularization in order to distinguish it from other techniques called ``spectral regularization".

Covariant spectral regularization is not associated with any form of divergence, renormalization is not required and one may obtain an expression for the QCD running coupling $\alpha_s(\tau)$ as a function of energy $\tau$ which does not require the setting of a renormalization scale $\mu_r$.

Peskin and Schroeder \cite{Peskin} in Chapter 6.2 on radiative corrections write in paragraph 2 on page 185 that, for QED, the vertex correction and quark self-energy diagrams represent corrections to an electron's response to a given applied field while the vacuum polarization diagram should be considered to be a correction to the electromagnetic field itself.
We consider the analogue of this statement to be the case for QCD when analyzed using spectral regularization. In other words we consider the QCD coupling to be determined by the dressed gluon propagator.

Use of the renormalization method in QCD generally requires counterterms which must be obtained from a number of diagrams such as the vacuum polarization, the electron self-energy and the vertex correction diagrams. 
If one uses the pinch technique in combination with the background field method (PT+BFM) \cite{Binosi} then the gluon polarization function captures all the required features of the renormalization group and hence the high energy behavior of the QCD running coupling. 

If covariant spectral regularization is used then, since renormalization is not involved, there are no  counterterms or the setting of a renormalization scale, the renormalization group equation method is not applicable and, we propose, the QCD running coupling can be determined from (that is, naturally defined to be obtained from) the behavior of the dressed gluon propagator (i.e. from vacuum polarization). We will call the QCD running coupling obtained in this manner (in particular the one-loop case) the spectral QCD running coupling and will, in this paper, be considering its properties.

We will present a method for determining the spectral QCD running coupling based on the relationship between the tree level Feynman amplitude ${\mathcal M}^{(\text{tree})}$ and a total Feynman amplitude ${\mathcal M}={\mathcal M}^{(\text{tree})}+{\mathcal M}^{(\text{vp})}$ made up of the tree level and the one-loop spectral vacuum polarization Feynman amplitudes for the $u\overline{d}\rightarrow u\overline{d}$ quark scattering process.

Our computations apply at all energies. There is no distinction between perturbative domain and non-perturbative domain. No input of experimental or lattice simulation data (other than the quark masses) is required.

\section{A brief review of previous work related to the QCD running coupling and the Landau pole}

The classical and most common approach to computing the QCD running coupling is to use the renormalization group equations as initially described by Stueckelberg and Petermann \cite{Stueckelberg} and further elucidated by Gell-Mann and Low \cite{Gell-Mann}. In this work the concept of the $\beta$-function, which is a function defined in terms of the running coupling function, was introduced. The renormalization group equations are applicable in the context of the renormalization method for dealing with infinities in Feynman integrals. These equations predict a Landau pole. For QED the energy at which this pole is located is extremely high while for QCD it is of the order of a few hundred MeV. Since a strong force Landau pole at the order of a few hundred MeV is unphysical, its prediction is commonly taken to indicate a breakdown of perturbation theory for QCD at low energies.

Peris and Rafael \cite{Peris} consider the type of singularities in Feynman series called renormalons which were proposed by 't Hooft in 1977 \cite{t_Hooft} and discussed in detail in Ref. \citen{Beneke}. They describe how renormalon singularities are often stated in the literature to be due to the Landau pole in the running coupling constant and how this belief is based on consideration of the first term of the $\beta$-function. They argue that renormalon singularities are not induced by a putative Landau pole when a two-loop $\beta$-function is employed.

Tissier and Wschebor \cite{Tissier} present a modification of the Fadeev-Popov (FP) action in the Landau gauge based on phenomenological considerations. They add a mass term for the gluon in the FP action. Their model exhibits no Landau pole and the running coupling $g_s$ is significantly smaller than 1 at all energies. As they mention, it is not possible to calculate the mass parameter in their theory from first principles.

Van Acoleyen and Verschelde \cite{VanAcoleyen} propose an alternative perturbative expansion for QCD in which all scheme and scale dependence is reduced to one free parameter and when the parameter is fixed using a  fastest apparent convergence criterion the resulting running coupling gives sensible results at all energies and does not diverge as one approaches the position of the Landau pole.

Arbuzov and Zaitsev \cite{Arbuzov} apply the so called Bogoliubov compensation principle to QCD. Using the principle (which they acknowledge is an approximation) they obtain a running coupling for QCD which does not manifest a Landau pole (however it does not have the property of ``freezing" in the infrared (see Section \ref{section:freezing} below)).

Maiezza and Vasquez \cite{Maiezza} argue, without reference to diagrammatic calculations, that under certain circumstances QFT must have a non-perturbative completion valid at all energies, and thus without a Landau pole.

Binosi {\em et al.} \cite{Binosi} describe a process independent running coupling ${\chp\alpha}_{\text{PI}}$ defined by interlacing two functions, one defined in the perturbative domain $k^2\gg m_0^2$ and the other defined in the non-perturbative domain $k^2\ll m_0^2$ where $m_0\approx m_p/2$ ($m_p=$ mass of the proton) involving interpolation of renormalization group continuum and/or lattice QCD calculations using a Pad\'{e} approximant. ${\chp\alpha}_{\text{PI}}$ coincides closely at low energy with the well known effective charge candidate $\alpha_{g_1}$ defined via the Bjorken sum rule in terms of proton and neutron structure functions which are determined experimentally.

de T\'{e}ramond {\em et al. \cite{deTeramond}} describe a model ostensibly related to holographic light front quantum chromodynamics (HLFQCD) which is a non-perturbative approach to QCD based on the AdS/CFT correspondence and light front quantization. They build a model in which they demand that in the deep infrared and ultraviolet domains the $\beta$-function vanishes. They model the effective running coupling, $\alpha_{\text{eff}}$, by the formula
\[ \alpha_{\text{eff}}(Q^2)=\alpha_{\text{eff}}(0)\exp\left[-\int_0^{Q^2}\frac{du}{4\kappa^2+u\text{ln}(\frac{u}{\Lambda^2})}\right], \]
for constants $\alpha_{\text{eff}}(0),\kappa$ and $\Lambda$, which they state follows from general considerations of symmetry and analyticity. They state that the above form for $\alpha_{\text{eff}}$ is valid in the non-perturbative and near perturbative domains but it remains an approximation in the deep ultraviolet fully perturbative region. They find that by suitably choosing the constants $\alpha_{\text{eff}}(0),\kappa$ and $\Lambda$ their model does indeed fit quite well the $(\overline{\text{MS}}$) experimental data in the domain of its applicability. Their model does not entirely banish the Landau pole but leads to its splitting into two complex conjugate singularities.

\section{Covariant Spectral Regularization\label{section:cov_spec_reg}}

Covariant spectral regularization has been applied to all the (one-loop) classical tests of QED radiative corrections and produces the same results as Pauli-Villars and dimensional regularization/renormalization, (though the technique is applicable to cases where the Feynman diagram has an arbitrary number of loops).

Covariant spectral regularization is based on the technique described in Ref.~\citen{Springer} and the use of the spectral calculus described in Ref.~\citen{Symmetry}.

In Ref. \citen{NPB} we use covariant spectral regularization to compute the QED vacuum polarization tensor and then use this to compute the Uehling potential function and the Uehling contribution to the Lamb shift for the hydrogen atom. 

In Ref. \citen{vertex} we use covariant spectral regularization to compute the vertex function for QED at arbitrary (on shell) values of its arguments in the t channel and the s channel. The low energy and low momenta limit of this function in the t channel is used to compute the one-loop contribution to the anomalous magnetic moment of the electron \cite{MSQUARE} and the result agrees with Schwinger's classical result. We also use the vertex function in the s channel to compute the vertex correction contribution (next-to-leading order (NLO) contribution) to the high energy limit of the cross section for the process $e^{+}e^{-}\rightarrow\gamma\rightarrow\mu^{+}\mu^{-}$, that is, muon production through electron-positron annihilation, and the result agrees with the textbook result for the cross section for this process. \cite{Schwartz} In the textbook computation of this cross section renormalization is used to cancel ultraviolet divergence and soft photon final state radiation is used to cancel infrared divergence. Our computation of this cross section does not use renormalization to cancel ultraviolet divergence or soft photon final state radiation to cancel infrared divergence, since there are no divergences, neither ultraviolet nor infrared.

\begin{figure} 
\centering
\includegraphics[width=3.5cm]{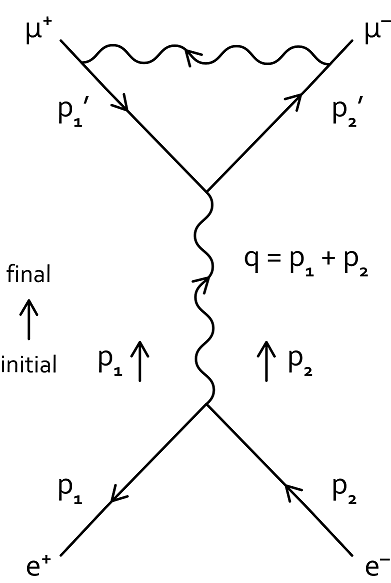}
\caption{Feynman diagram for NLO contribution to muon production\label{fig:Fig1}}
\end{figure}

Thus our computation of the NLO contribution for muon production is associated with the Feynman diagram of Fig.~\ref{fig:Fig1}, no final state radiation is involved. The NLO contribution for muon production with final state radiation can be computed by considering the Feynman diagram of Fig.~\ref{fig:Fig2} and using covariant spectral regularization.

\begin{figure} 
\centering
\includegraphics[width=3cm]{Fig2.png}
\caption{NLO contribution to muon production with final state radiation\label{fig:Fig2}}
\end{figure}

Therefore we have used covariant spectral regularization on all the classical tests of QED radiative corrections and our results agree with the classical results. Specifically, the references describing these checks are as follows
\begin{itemize}
\item Uehling contribution to the Lamb shift - Refs. \citen{IJMPA,NPB}
\item  anomalous magnetic moment of the electron - Ref. \citen{MSQUARE}
\item $e^{+}e^{-}\rightarrow\gamma\rightarrow\mu^{+}\mu^{-}$ - Ref. \citen{vertex}
\item $e^{-}$ self energy - Ref. \citen{Lamb_shift}
\end{itemize}
While the classical calculations use complicated arguments to negotiate divergences, covariant spectral regularization does not involve divergence of any sort.

In the present paper we use covariant spectral regularization to compute the spectral QCD vacuum polarization tensor and then use this to compute the spectral QCD running coupling.

\section{Feynman diagrams, Feynman amplitudes and vacuum polarization tensors}

Feynman diagrams in general are labeled directed graphs (di-graphs) whose edges may be of type electron, $\mu$, $\tau$, their corresponding neutrinos, various quark types, their anti-particles, photon, Z$^{0}$, W$^{\pm}$, gluon (or Higgs).

We consider pure QCD Feynman diagrams whose edges may be of type gluon or quark. We are interested in the process $u\overline{d}\rightarrow u\overline{d}$ of up quark anti-down quark scattering.

In the standard model of particle physics \cite{Schwartz} physical quantities of interest about this process (specifically its differential cross section) can be obtained from the object called the Feynman amplitude for the process which can be obtained by summing all (or some) of the individual amplitudes associated with Feynman graphs with inputs of type $u,\overline{d}$ and outputs of type $u,\overline{d}$ (such that a continuous sequence of edges of type $u$ may be traced from the input $u$ edge to the output $u$ edge, similarly for the $\overline{d}$ edges). The resulting series is generally asymptotic, hence divergent, even after divergences of individual terms have been dealt with by renormalization. The overall divergence of the asymptotic series may be dealt with by considering a finite sub-series or else by using techniques such as Borel summation.

\begin{figure} 
\centering
\includegraphics[width=11cm]{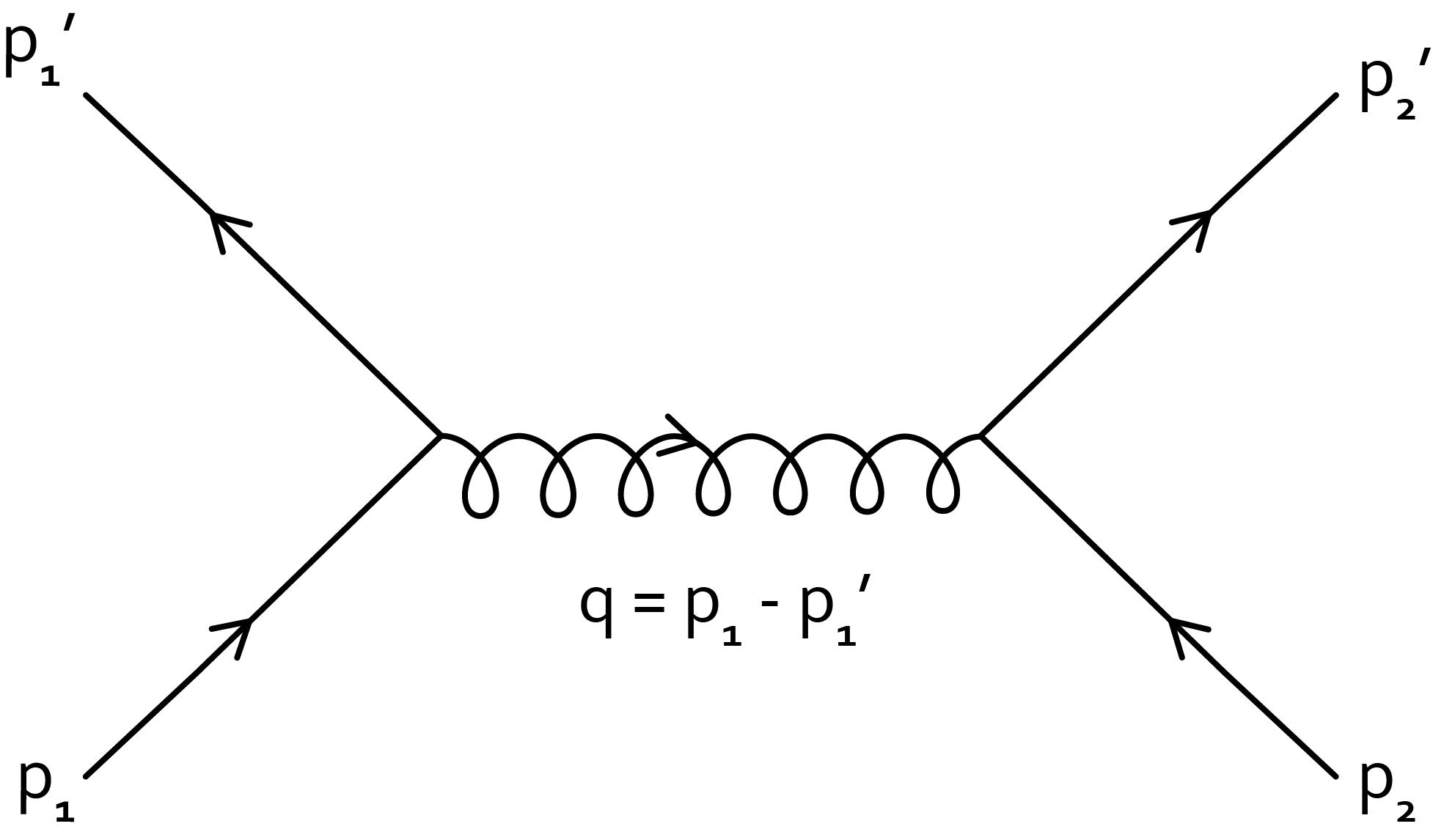}
\caption{Tree level quark $u\overline{d}$ scattering\label{fig:tree_level}}
\end{figure}

\begin{figure} 
\centering
\includegraphics[width=12cm]{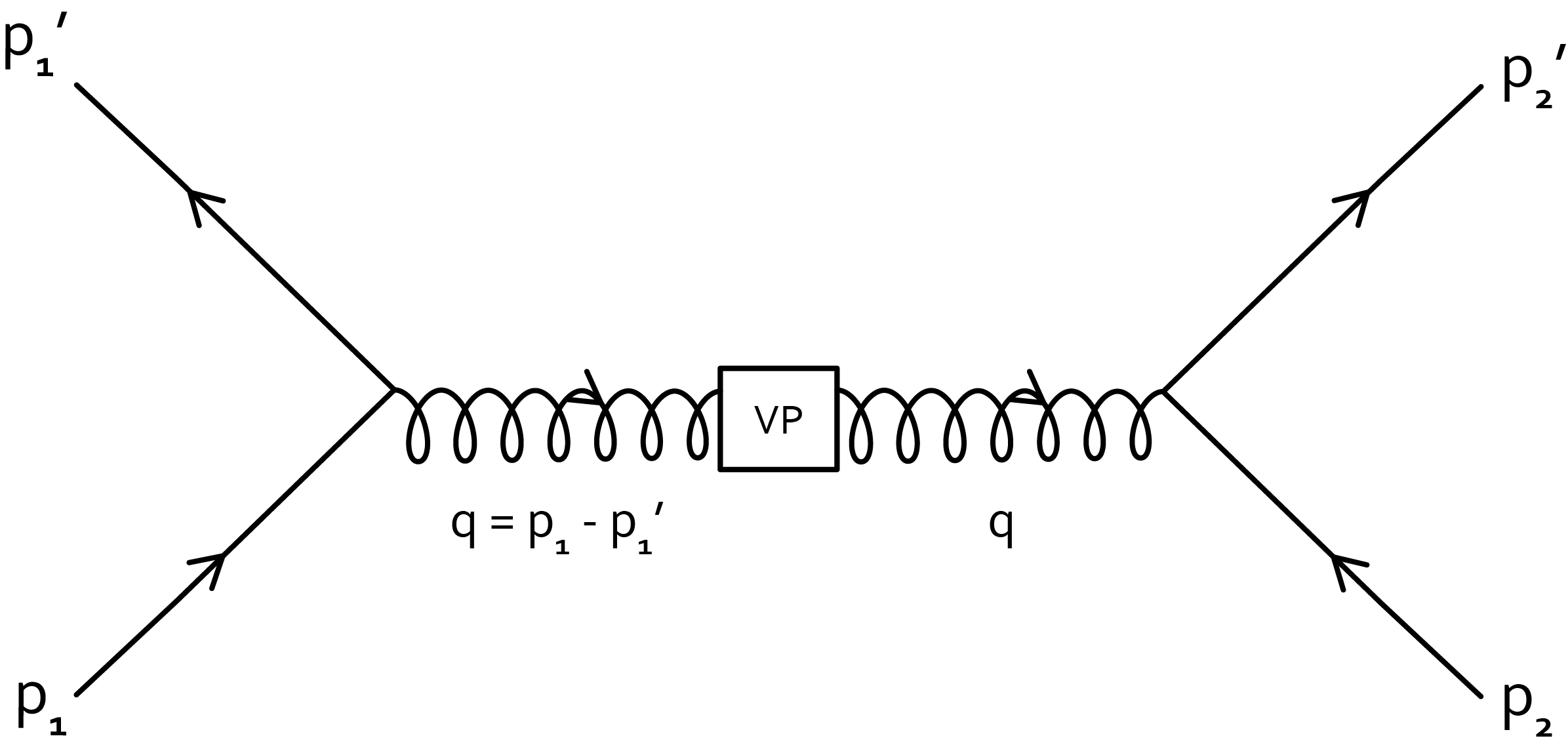}
\caption{Vacuum polarization for quark $u\overline{d}$ scattering\label{fig:vacuum_polarization}}
\end{figure}

The individual Feynman amplitude for each component graph may be obtained by application of the Feynman rules \cite{Schwartz}. The simplest and predominant of such graphs is the so called tree level graph shown in Fig.~\ref{fig:tree_level} which forms an acyclic di-graph.
Another class of graphs for this process, shown in Fig.~\ref{fig:vacuum_polarization}, is obtained by the so called dressing of the gluon propagator of the tree level graph.

Such a dressed gluon propagator forms what is known as a vacuum polarization contribution. In general a QCD vacuum polarization graph is any Feynman graph which has exactly one input consisting of a gluon edge and exactly one output consisting of a gluon edge. In the present paper we will be concerned with one-loop vacuum polarization graphs.

\section{Spectral QCD Vacuum Polarization}

We construct the (one-loop) spectral QCD vacuum polarization tensor $\Pi^{\text{(vp)}}$ as
\[ \Pi^{\text{(vp)}}=\Pi^{(q)}+\Pi^{(g)}+\Pi^{(4)}, \]
where $\Pi^{(q)}$ is associated with the quark fermion bubble diagram, $\Pi^{(g)}$ is associated with the gluon bubble diagram and $\Pi^{(4)}$ is associated with the four-point gluon bubble diagram. 

We do not include a diagram for ``ghosts". On p. 508 of Ref. \citen{Schwartz} Schwartz writes ``That we need these ghosts is a horrible consequence of the Lagrangian formulation of field theory". Our formulation of field theory is not based on Lagrangians but rather on $U(2,2)$ covariance (in locally conformally flat space-time $X$) which gives rise to $U(1)\times SU(2)\times SU(3)$ covariance (in the tangent space $T_xX$) \cite{AMP,IJMPA,AIPAdv}. Therefore we do not ``believe in ghosts" in this context. 

\subsection{Brief outline of how quantum field theory can be described in terms of $U(2,2)$ covariance}
This approach is explained in detail in Ref.~\citen{IJMPA} and, especially, in Ref.~\citen{AIPAdv}. In this approach we model space-time $X$ as a locally conformally flat Lorentzian manifold (i.e., what is known as  a M\"{o}bius structure). Such objects are associated with a number of principal bundles, in particular a principal bundle with structure group $U(2,2)$. Such bundles are associated with infinitesimal Lie algebra bundles with structure algebra ${\mathfrak g}=u(2,2)$.

Let \[ A=\left(\begin{array}{cc}
a&b\\
c&d
\end{array}\right)\in gl(4,{\bf C})={\bf C}^{4\times4}, \text{ where } a,b,c,d\in{\bf C}^{2\times2}. \]
Then
\begin{align*} 
A\in u(2,2)&\Leftrightarrow A^{\dagger}=-gAg\\
&\Leftrightarrow\left(\begin{array}{cc}
a^{\dagger}&c^{\dagger}\\
b^{\dagger}&d^{\dagger}
\end{array}\right)=-\left(\begin{array}{cc}
0&1\\
1&0
\end{array}\right)\left(\begin{array}{cc}
a&b\\
c&d
\end{array}\right)\left(\begin{array}{cc}
0&1\\
1&0
\end{array}\right)=\left(\begin{array}{cc}
-d&-c\\
-b&-a
\end{array}\right)\\
&\Leftrightarrow c^{\dagger}=-c,b^{\dagger}=-b,d=-a^{\dagger}, \end{align*}
where 
\[ g=\left(\begin{array}{cc}
0&1\\1&0
\end{array}\right) (=\gamma^0 \text{ in the chiral representation for the Dirac gamma matrices}), \]
is the metric tensor for $U(2,2)$. Thus we have natural linear maps $\Xi_1:{\mathfrak g}\rightarrow u(2)$ and $\Xi_2:{\mathfrak g}\rightarrow gl(2,{\bf C})$ defined by
\[ \Xi_1(A)=b,\text{ and }\Xi_2(A)=a. \]

Letting $\Theta$ be any real linear isomorphism from $gl(2,{\bf C})$ to $su(3)$ (both being 8 dimensional) we have associated with $X$ a $u(2)\cong u(1)\times su(2)$ Lie algebra bundle and a $su(3)$ Lie algebra bundle (though $\Theta$ is not a Lie algebra isomorphism). These are associated with $U(1)\times SU(2)$ and $SU(3)$ principal bundles.

It is shown in Ref.~\citen{AIPAdv} (and one could say that it is quite well known) how the electroweak interaction can be understood in terms of $U(1)\times SU(2)$ covariance. Also, the strong interaction can be understood in terms of $SU(3)$ covariance (the 8-fold way).

\subsection{Method of construction of $\Pi^{(q)}$ and $\Pi^{(g)}$}
 
We now proceed to compute $\Pi^{(q)}$ and $\Pi^{(g)}$ using spectral regularization. In these computations $\Pi^{(q)}$ and $\Pi^{(g)}$ are first represented as (Lorentz covariant) tensor valued measures on Minkowski space (which is isomorphic to $T_xX$) and then spectral calculus \cite{Symmetry} is used to compute densities which exactly generate these measures (with respect to Lebesgue measure on Minkowski space). These densities are functions which can be used for $\Pi^{(q)}$ and $\Pi^{(g)}$ in calculations involving $\Pi^{\text{(vp)}}$.

We also show that $\Pi^{(4)}$ vanishes.

\section{The quark fermion bubble\label{section:quark_bubble}}

\begin{figure} 
\centering
\includegraphics[width=10cm]{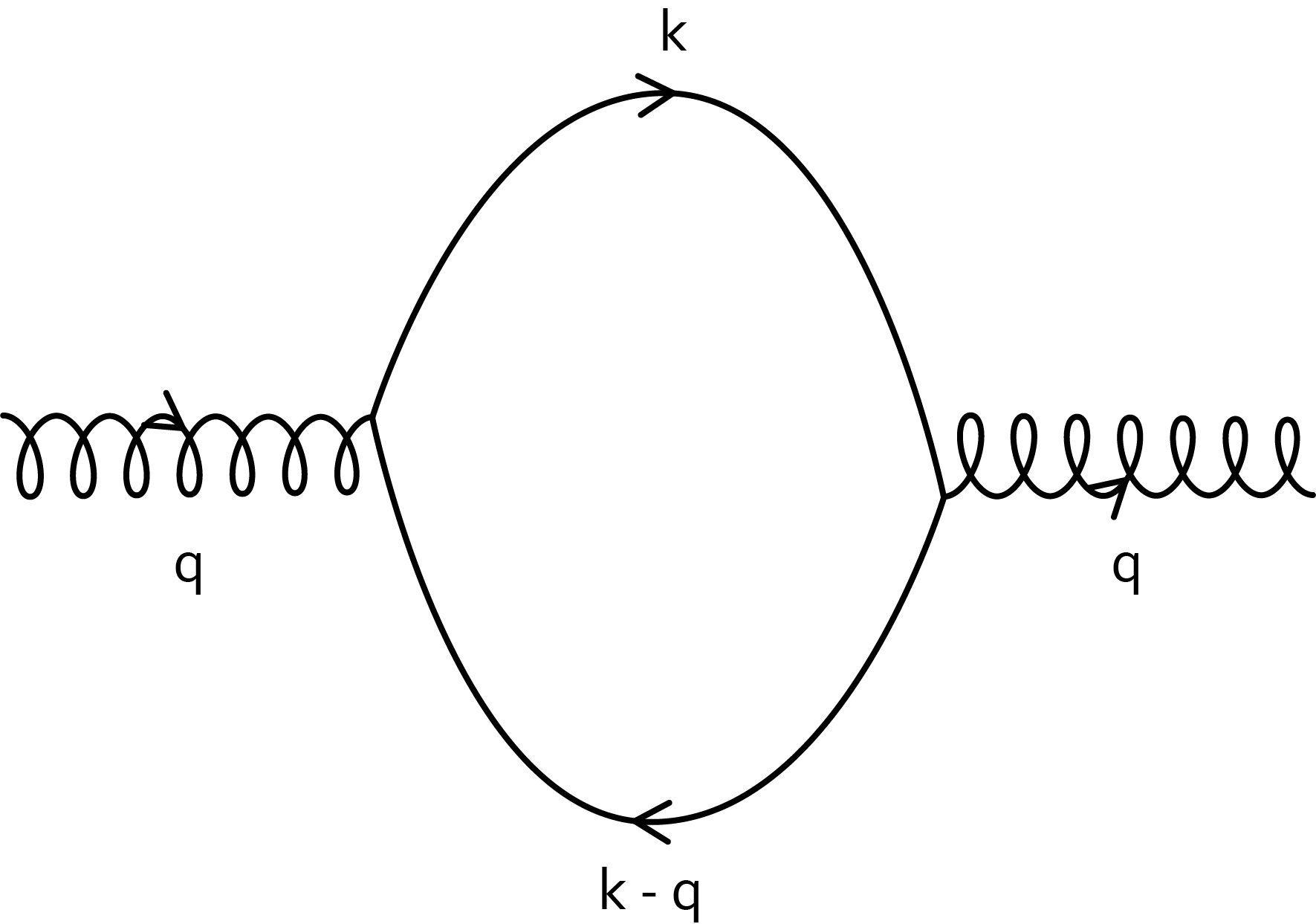}
\caption{Quark bubble\label{fig:Qbubble}}
\end{figure}

\begin{figure} 
\centering
\includegraphics[width=10cm]{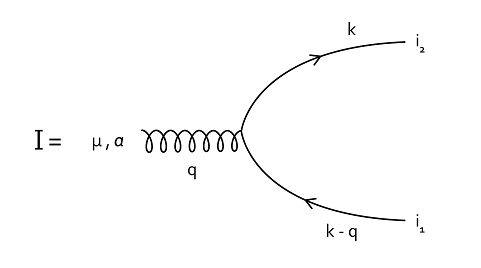}
\caption{Quark bubble contribution I\label{fig:quark_bubble_I}}
\end{figure}

\begin{figure} 
\centering
\includegraphics[width=10cm]{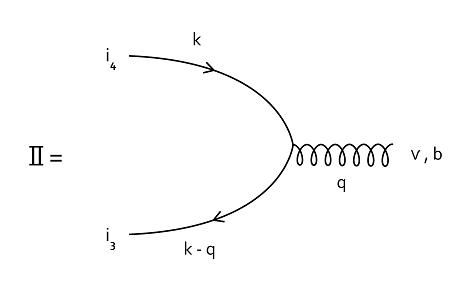}
\caption{Quark bubble contribution II\label{fig:quark_bubble_II}}
\end{figure}

The quark fermion bubble is shown in Fig.~\ref{fig:Qbubble}.
This diagram can be considered to be generated by combining the two diagrams, I and II shown in Figs. \ref{fig:quark_bubble_I} and \ref{fig:quark_bubble_II}.
 Using the Feynman rules \cite{Schwartz} these sub-graphs are associated with the quantities
\begin{align*}
I^{a\mu}_{i_1i_2}=&g_s\gamma^{\mu}T^{a}_{i_1i_2},\\
II^{b\nu}_{i_3i_4}=&g_s\gamma^{\nu}T^{b}_{i_3i_4},
\end{align*}
where $g_s$ is the strong coupling constant, $\{\gamma^{\mu}\}_{\mu=0}^3$ are the Dirac gamma matrices and $\{T^{a}_{ij}\}_{a=1,\ldots,8;i,j=1,2,3}$ are the generators for the Lie algebra $su(3)$ for the gauge group $SU(3)$.

To compute the Feynman amplitude associated with the quark bubble, following the Feynman rules, we first compute the quantity $\Psi^{ab\mu\nu}$ in the space
\[ \text{(gluon color space)} \otimes \text{(quark configuration space)}, \]
defined by
\begin{align}
\Psi^{ab\mu\nu}_{ij\alpha\beta}(q)=-\int\frac{dk}{(2\pi)^4}\left(iI^{a\mu}_{ ii_1}iS_{\text{QCD},i_1i_2}(k-q)iII^{b\nu}_{i_2i_3}iS_{\text{QCD},i_3j}(k)\right)_{\alpha\beta},\label{eq:Psi_def}
\end{align}
summing over repeated gluon color indices (Einstein summation convention), where $S_{\text{QCD}}$ is the QCD fermion propagator defined by
\[ S_{\text{QCD},ij}(p)=\frac{\delta{ij}}{{\slas p}-m+i\epsilon}, \]
(the minus sign in Eq.~\ref{eq:Psi_def} occurs because the quark bubble is a fermion loop). Thus, since
\[ \frac{1}{{\slas p}-m+i\epsilon}=\frac{{\slas p}+m}{p^2-m^2+i\epsilon}, \]
we have that
\begin{align}
&\Psi^{ab\mu\nu}_{ij}(q)\nonumber\\
=&-\int\frac{dk}{(2\pi)^4}g_s\gamma^{\mu}T^{a}_{ii_1}\frac{\delta_{i_1i_2}({\slas k}-{\slas q}+m)}{(k-q)^2-m^2+i\epsilon}g_s\gamma^{\nu}T^{b}_{i_2i_3}\frac{\delta_{i_3j}({\slas k}+m)}{k^2-m^2+i\epsilon}\nonumber\\
=&-g_s^2(T^{a}T^{b})_{ij}\int\frac{dk}{(2\pi)^4}\gamma^{\mu}({\slas k}-{\slas q}+m)\gamma^{\nu}({\slas k}+m)\frac{1}{(k-q)^2-m^2+i\epsilon}\nonumber\\
&\frac{1}{k^2-m^2+i\epsilon},\label{eq:Psi_implicit}
\end{align}
where the quark configuration space indices in Eq.~\ref{eq:Psi_implicit} are implicit, i.e. it is a matrix equation in ${\bf C}^{4\times4}$.
Finally to compute the Feynman amplitude $\Pi^{(q)ab\mu\nu}$ of the quark bubble we take the trace Tr in (gluon color space) $\otimes$ (quark configuration space) of $\Psi^{ab\mu\nu}$, defined by
\[ \Pi^{(q)ab\mu\nu}(q)=\text{Tr}\left(\Psi^{ab\mu\nu}(q)\right)=\sum_{i=1}^3\sum_{\alpha=0}^3\left(\Psi^{ab\mu\nu}_{ii\alpha\alpha}(q)\right)=\sum_{i=1}^3\left(\text{Tr}\left(\Psi^{ab\mu\nu}_{ii}(q)\right)\right). \]   
 
Therefore the contribution of the quark bubble to the QCD vacuum polarization is 
\begin{align*}
\Pi^{(q)ab\mu\nu}(q)&=-g_s^2\text{Tr}(T^aT^b)\int\frac{dk}{(2\pi)^4}\frac{1}{(q-k)^2-m^2+i\epsilon}\frac{1}{k^2-m^2+i\epsilon}\\
& \text{Tr}[\gamma^{\mu}({\slas k}-{\slas q}+m)\gamma^{\nu}({\slas k}+m)].
\end{align*}
Thus, since
\[ \text{Tr}(T^aT^b)=\frac{1}{2}\delta^{ab}, \]
we have
\begin{equation}\label{eq:full_Pi_def}
\Pi^{(q)ab\mu\nu}(q)=\frac{1}{2}g_s^2\delta^{ab}\Pi^{(q)\mu\nu}(q),
\end{equation}
where
\begin{align}\label{eq:divergent_integral}
\Pi^{(q)\mu\nu}(q)=&-\int\frac{dk}{(2\pi)^4}\frac{1}{(q-k)^2-m^2+i\epsilon}\frac{1}{k^2-m^2+i\epsilon} \nonumber\text{Tr}[\gamma^{\mu}({\slas k}-{\slas q}+m)\gamma^{\nu}\nonumber\\
&({\slas k}+m)].
\end{align}
Unfortunately, as is well known, the integral given by Eq.~\ref{eq:divergent_integral} is divergent for any given $q\in{\bf R}^4$. Applying the first step in the general method of spectral regularization (as in the work of Refs. \citen{IJMPA,NPB}) suppose  (``pretend") that $\Pi^{(q)\mu\nu}$ existed pointwise and then we can compute (formally) the measure associated with the ``function"
\begin{align*}
q\mapsto\Pi^{\mu\nu}(q)=&-\int\frac{1}{(q-k)^2-m^2+i\epsilon}\frac{1}{k^2-m^2+i\epsilon}\text{Tr}[\gamma^{\mu}({\slas k}-{\slas q}+m)\gamma^{\nu}\\
&({\slas k}+m)]\,dk,
\end{align*}
as follows. Let ${\mathcal B}_0({\bf R}^4)$ denote the set of relatively compact Borel subsets of ${\bf R}^4$. Now let $\Gamma\in{\mathcal B}_0({\bf R}^4)$. Then
\begin{align}
\Pi^{\mu\nu}(\Gamma)&=\int_{q\in\Gamma}\Pi^{\mu\nu}(q)\,dq\nonumber\\
&=\int\chi_{\Gamma}(q)\Pi^{\mu\nu}(q)\,dq\nonumber\\
&=-\int\chi_{\Gamma}(q)\frac{1}{(q-k)^2-m^2+i\epsilon}\frac{1}{k^2-m^2+i\epsilon}\nonumber\\
&\text{Tr}[\gamma^{\mu}({\slas k}-{\slas q}+m)\gamma^{\nu}({\slas k}+m)]\,dk\,dq\nonumber\\
&=-\int\chi_{\Gamma}(q)\frac{1}{(q-k)^2-m^2+i\epsilon}\frac{1}{k^2-m^2+i\epsilon}\nonumber\\
&\text{Tr}[\gamma^{\mu}({\slas k}-{\slas q}+m)\gamma^{\nu}({\slas k}+m)]\,dq\,dk\nonumber\\
&=\int\chi_{\Gamma}(q+k)\frac{1}{q^2-m^2+i\epsilon}\frac{1}{k^2-m^2+i\epsilon}\nonumber\\
&\text{Tr}[\gamma^{\mu}({\slas q}-m)\gamma^{\nu}({\slas k}+m)]\,dq\,dk\nonumber\\
&=-\pi^2\int\chi_{\Gamma}(q+k)\text{Tr}[\gamma^{\mu}({\slas q}-m)\gamma^{\nu}({\slas k}+m)]\,\Omega_m(dq)\Omega_m(dk),\label{sequence_of_steps}
\end{align}
where, for any set $\Gamma$, $\chi_{\Gamma}$ denotes the characteristic function of $\Gamma$ defined by
\[ \chi_{\Gamma}(q)=\left\{\begin{array}{l}
1\text{ if }q\in\Gamma,\\
0\text{ otherwise,}
\end{array}\right. \]
and we have used the result \cite{Symmetry,IJMPA,NPB,AIPAdv} that, for on shell $q$,
\[ \frac{1}{q^2-m^2+i\epsilon}\rightarrow-\pi i\Omega_m^{\pm}(q), \]
in which, for $m\geq0$, $\Omega_m^{\pm}$ denotes the standard Lorentz invariant measure on ${\bf R}^4$ concentrated on the mass shell $H_m^{\pm}=\pm\{q\in{\bf R}^4:q^2=m^2,q^0>0\}$ given by
\[ \Omega_m^{\pm}(\Gamma)=\int_{{\vct q}\in{\bf R}^3}\chi_{\Gamma}(\pm\omega_m({\vct q}),{\vct q})\omega_m({\vct q})^{-1}\,d{\vct q}, \]
where $\omega_m({\vct q})=(m^2+{\vct q}^2)^{\frac{1}{2}}$ and $\Omega_m=\Omega_m^{+}$.
\cite{Symmetry,IJMPA,NPB}

Therefore the measure associated with $\Pi^{(q)\mu\nu}$ is
\[ \Pi^{(q)\mu\nu}(\Gamma)=-\frac{1}{16\pi^2}\int\chi_{\Gamma}(q+k)\text{Tr}[\gamma^{\mu}({\slas q}-m)\gamma^{\nu}({\slas k}+m)]\,\Omega_m(dq)\Omega_m(dk). \]
The important thing is that, while the integral given by Eq.~\ref{eq:divergent_integral}  is divergent for $q\in{\bf R}^4$, the measure $\Pi^{(q)\mu\nu}$ that we have associated with it is a well defined measure on ${\bf R}^4$.

$\Pi^{(q)\mu\nu}$ is analogous to the measure associated with the QED vacuum polarization tensor and it can be shown \cite{NPB,IJMPA} that $\Pi^{(q)\mu\nu}$ is a causal tempered Borel measure on Minkowski space and is associated with a density $\Pi^{(q)\mu\nu}$ given by
\begin{equation*}
\Pi^{(q)\mu\nu}(q)=(q^2\eta^{\mu\nu}-q^{\mu}q^{\nu})\pi^{(q)}(s)=(s^2\eta^{\mu\nu}-q^{\mu}q^{\nu})\pi^{(q)}(s),
\end{equation*}
where
\begin{equation}\label{eq:pi_q_def}
\pi^{(q)}(s)=-\frac{2}{3\pi}m^3s^{-3}Z(s)(3+2Z(s)^2),
\end{equation}
$s=(q^2)^{\frac{1}{2}}$ and 
\[ Z(s)=\left\{\begin{array}{ll}
(\frac{s^2}{4m^2}-1)^{\frac{1}{2}}&\text{ if }s\geq2m,\\
0&\text{ otherwise.}
\end{array}\right. \]
The reader may wonder why the integral given by Eq.~\ref{eq:divergent_integral} is divergent so the function that it purports to define is not well defined, yet we claim to obtain, by a rigorous argument, a well defined function $\Pi^{(q)\mu\nu}$ associated with it. The answer is that our argument is rigorously correct at every step except line 4 of the sequence (\ref{sequence_of_steps}) of equations where we interchange the order of integration, which is not universally valid. 

The measure $\Gamma\mapsto\Pi^{(q)\mu\nu}(\Gamma)$ is causal \cite{Symmetry}. Therefore the function $q\mapsto\Pi^{(q)\mu\nu}(q)$ is defined in the timelike domain. We would like to compute $\Pi^{(q)\mu\nu}$ in the spacelike domain because we are considering a scattering process and, for such processes, the momentum transfer is spacelike. We do this by analytically continuing $\Pi^{(q)\mu\nu}$ from the timelike domain to the spacelike domain. 

In the spacelike domain we have, following \cite{NPB}, since (the analytic continuation of) $\pi^{(q)}$ (to the whole real line) is odd in $s$, that $\Pi^{(q)}$ is given by
\begin{align}
\Pi^{(q)\mu\nu}(q)=&(q^2\eta^{\mu\nu}-q^{\mu}q^{\nu})\pi^{(q)}(-Q)\nonumber\\
=&-(-Q^2\eta^{\mu\nu}-q^{\mu}q^{\nu})\pi^{(q)}(Q)\nonumber\\
=&(Q^2\eta^{\mu\nu}+q^{\mu}q^{\nu})\pi^{(q)}(Q),\label{eq:Pi_mu_nu}
\end{align}
where $Q=(-q^2)^{\frac{1}{2}}$. 
Therefore, by Eqns.~\ref{eq:full_Pi_def}, \ref{eq:pi_q_def} and \ref{eq:Pi_mu_nu},
\begin{align*}
\Pi^{(q)ab\mu\nu}(q)=&-\frac{1}{3\pi}g_s^2\delta^{ab}m^3Q^{-3}Z(Q)(3+2Z(Q)^2)(Q^2\eta^{\mu\nu}+q^{\mu}q^{\nu}),
\end{align*}
where $Q=(-q^2)^{\frac{1}{2}}$.

Now there are six quark types with flavors $f_q=u, d, s, c, b, t$ and masses $m_q\in\{m_1,\ldots,m_6\}$ say. Therefore we take the total quark bubble contribution to the hadronic vacuum polarization tensor to be given by
\begin{equation}\label{eq:total_vp}
\Pi^{(q,\text{tot})ab\mu\nu}(q)=\sum_{k=1}^6\Pi^{(q)ab\mu\nu}(m_k,q),
\end{equation}
where
\begin{align*}
&\Pi^{(q)ab\mu\nu}(m,q)=-\frac{1}{3\pi}g_s^2\delta^{ab}m^3Q^{-3}Z(m,Q)(3+2Z(m,Q)^2)(Q^2\eta^{\mu\nu}+q^{\mu}q^{\nu}),
\end{align*}
in which
\[ Z(m,Q)=\left\{\begin{array}{ll}
\left(\frac{Q^2}{4m^2}-1\right)^{\frac{1}{2}}&\text{ if }Q\geq2m,\\
0&\text{ otherwise,}\end{array}\right. \]
and $Q=(-q^2)^{\frac{1}{2}}$. 

It is not necessary to specify which quarks are ``active" at any given energy. This is automatically worked out in Eq.~\ref{eq:total_vp} by the value determined by that  equation corresponding to any given quark type at any given energy. 

$\Pi^{(q,\text{tot})}$ stands for the six different quark bubble diagrams that need to be considered when computing the total one-loop vacuum polarization.

\section{The gluon bubble\label{section:gluon_bubble}}

\begin{figure} 
\centering
\includegraphics[width=12cm]{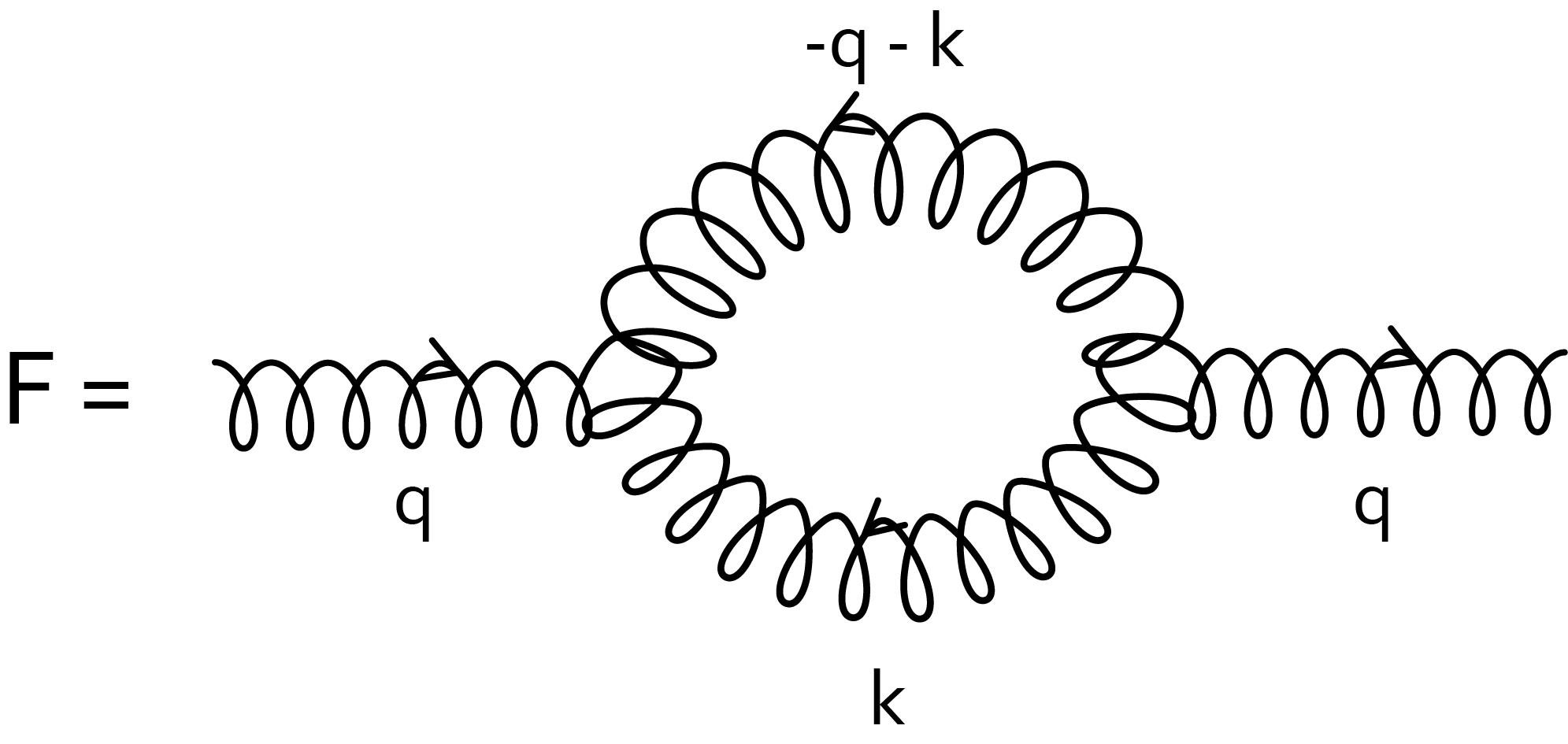}
\caption{gluon bubble diagram F\label{fig:gluon_bubble_F}}
\end{figure}

\begin{figure} 
\centering
\includegraphics[width=8cm]{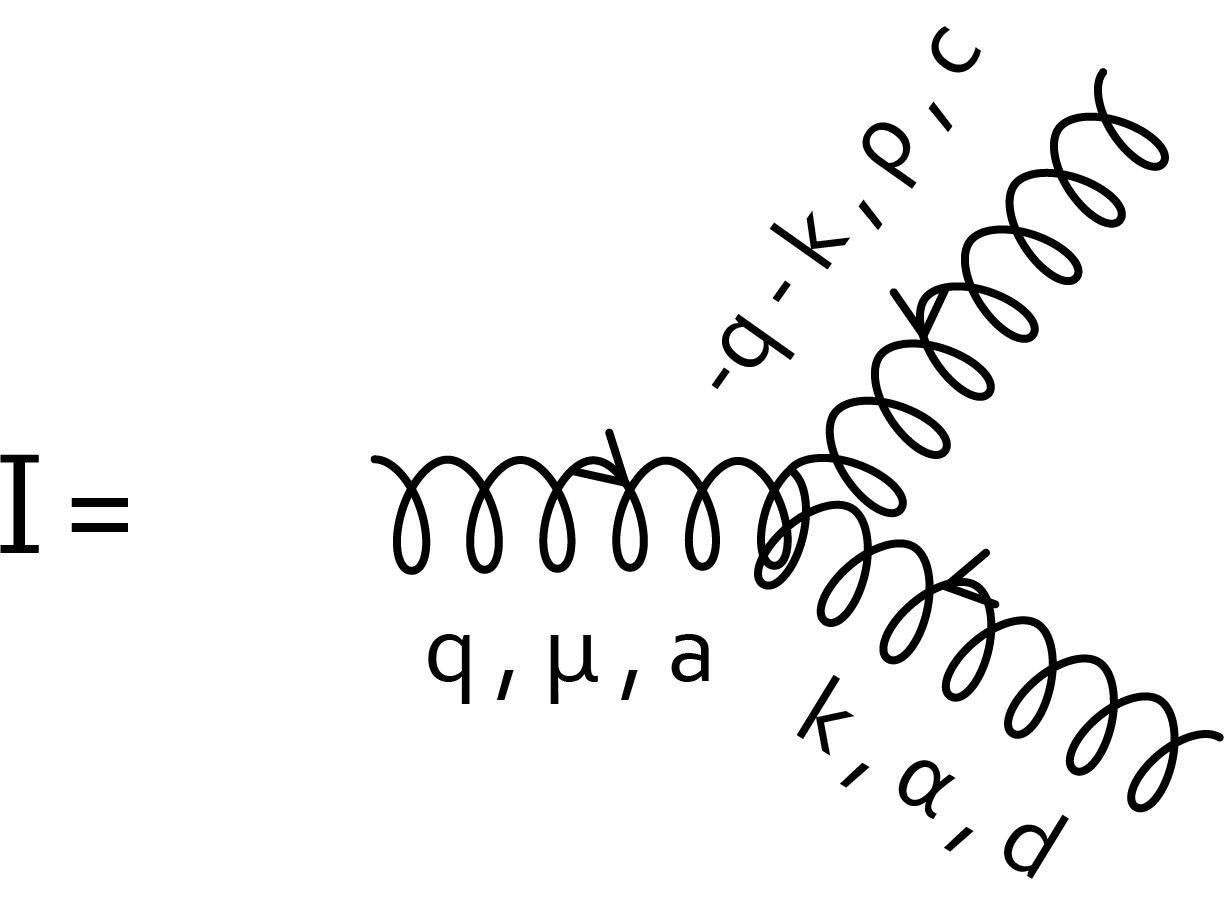}
\caption{gluon bubble diagram I\label{fig:gluon_bubble_I}}
\end{figure}

\begin{figure} 
\centering
\includegraphics[width=8cm]{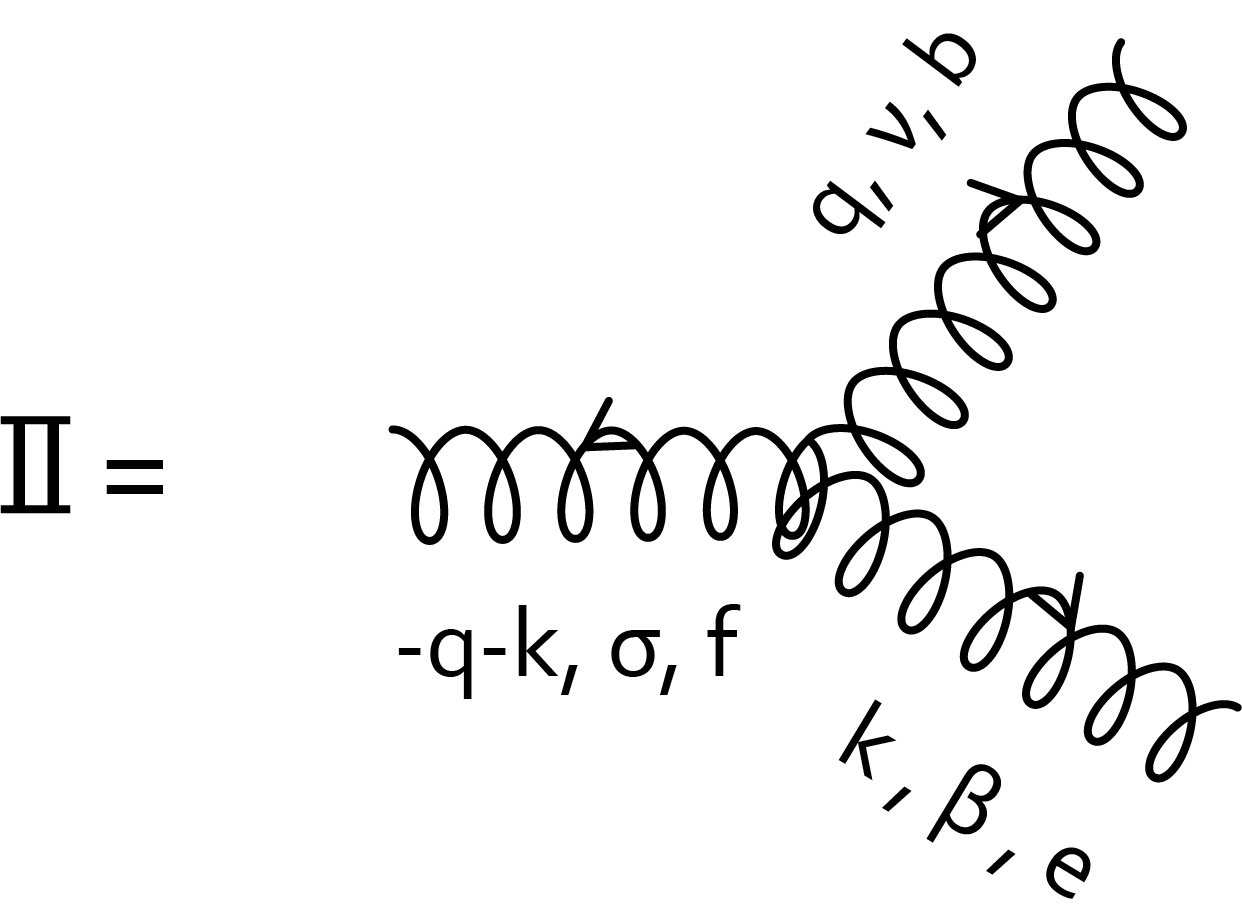}
\caption{gluon bubble diagram II\label{fig:gluon_bubble_II}}
\end{figure}

The contribution from the gluon bubble is 
\begin{eqnarray}
\Pi^{(g)ab\mu\nu}(q) & = & \frac{1}{2}\int\frac{dk}{(2\pi)^4}\frac{-i}{k^2+i\epsilon}\frac{-i}{(k+q)^2+i\epsilon}F^{ab\mu\nu}(q,k) \nonumber \\
& =  & -\frac{1}{2}\int\frac{dk}{(2\pi)^4}\frac{1}{k^2+i\epsilon}\frac{1}{(k+q)^2+i\epsilon}F^{ab\mu\nu}(q,k),\label{eq:gluon_bubble}
\end{eqnarray}
where $F^{ab\mu\nu}$ is defined by the Feynman diagram F shown in Fig.~\ref{fig:gluon_bubble_F}.
F is constructed from the diagrams I and II shown in Figs.~\ref{fig:gluon_bubble_I} and \ref{fig:gluon_bubble_II}.


The overall factor of $\frac{1}{2}$ is a symmetry factor required because the gluons are their own antiparticle (see p. 518 of Ref. \citen{Schwartz}).

Using the QCD Feynman rules (see p. 510 of Ref. \citen{Schwartz}) we have
\begin{align*}
I^{cad\rho\mu\alpha}(q,k)=&g_sf^{cad}(\eta^{\rho\mu}(-2q-k)^{\alpha}+\eta^{\mu\alpha}(q-k)^{\rho}+\eta^{\alpha\rho}(2k+q)^{\mu}),\\
II^{bfe\nu\sigma\beta}(q,k)=&g_sf^{bfe}(\eta^{\nu\sigma}(-2q-k)^{\beta}+\eta^{\sigma\beta}(2k+q)^{\nu}+\eta^{\beta\nu}(q-k)^{\sigma},
\end{align*}
and
\[ F^{ab\mu\nu}(q,k)=\delta^{cf}\delta^{de}\eta^{\rho\sigma}\eta^{\alpha\beta}I^{cad\rho\mu\alpha}(q,k)II^{bfe\nu\sigma\beta}(q,k). \]
Now
\[ \delta^{cf}\delta^{de}g_sf^{cad}g_sf^{bfe}=g_s^2f^{cad}f^{bcd}=-g_s^2f^{acd}f^{bcd}=-3g_s^2\delta^{ab}. \]
Therefore
\[ F^{ab\mu\nu}(q,k)=-3g_s^2\delta^{ab}N^{\mu\nu}(q,k), \]
where
\begin{eqnarray}
N^{\mu\nu}(q,k)  & = & [\eta^{\mu\alpha}(q-k)^{\rho}+\eta^{\alpha\rho}(q+2k)^{\mu}-\eta^{\rho\mu}(k+2q)^{\alpha}]\eta_{\alpha\beta}\eta_{\rho\sigma} \nonumber \\
    &  & \times[\eta^{\nu\beta}(q-k)^{\sigma}+\eta^{\beta\sigma}(2k+q)^{\nu}-\eta^{\sigma\nu}(2q+k)^{\beta}], \nonumber
\end{eqnarray}
and so, by Eq.~\ref{eq:gluon_bubble},
\begin{equation}\label{eq:gluon_bubble_1}
\Pi^{(g)ab\mu\nu}=\frac{3}{32\pi^4}g_s^2\delta^{ab}\Xi^{\mu\nu}(q)\,dq,
\end{equation}
where
\[ \Xi^{\mu\nu}(q)=\int\frac{1}{k^2+i\epsilon}\frac{1}{(k+q)^2+i\epsilon}N^{\mu\nu}(q,k)\,dk. \]
We compute (formally) the measure $\Xi^{\mu\nu}$ associated with the ``function"  $q\mapsto\Xi^{\mu\nu}(q)$ as follows. 
\begin{align}
\Xi^{\mu\nu}(\Gamma)&=\int_{q\in\Gamma}\Xi^{\mu\nu}(q)\,dq\nonumber \\
&=\int\chi_{\Gamma}(q)\,\Xi^{\mu\nu}(q)\,dq\nonumber\\
&=\int\chi_{\Gamma}(q)\frac{1}{k^2+i\epsilon}\frac{1}{(k+q)^2+i\epsilon}N^{\mu\nu}(q,k)\,dk\,dq\nonumber\\
&=\int\chi_{\Gamma}(q)\frac{1}{k^2+i\epsilon}\frac{1}{(k+q)^2+i\epsilon}N^{\mu\nu}(q,k)\,dq\,dk\nonumber\\
&=\int\chi_{\Gamma}(q-k)\frac{1}{k^2+i\epsilon}\frac{1}{q^2+i\epsilon}N^{\mu\nu}(q-k,k)\,dq\,dk\nonumber\\
&=\int\chi_{\Gamma}(q+k)\frac{1}{k^2+i\epsilon}\frac{1}{q^2+i\epsilon}N^{\mu\nu}(q+k,-k)\,dq\,dk\nonumber\\
&=-\pi^2\int\chi_{\Gamma}(q+k)N^{\mu\nu}(q+k,-k)\,\Omega_0(dq)\,\Omega_0(dk).\label{eq:Xi_steps} 
\end{align}
Once again the only step in the argument which is not rigorously correct is the interchange of the order of integration which occurs at line 4 of the sequence (\ref{eq:Xi_steps}) of equations.

For each $\mu,\nu\in\{0,1,2,3\}$ define $M^{\mu\nu}:{\bf R}^4\times{\bf R}^4\rightarrow{\bf R}$ by
\[ M^{\mu\nu}(q,k)=N^{\mu\nu}(q+k,-k). \]
Then
\begin{equation}\label{eq:Xi}
\Xi^{\mu\nu}(\Gamma)=-\pi^2\int\chi_{\Gamma}(q+k)M^{\mu\nu}(q,k)\,\Omega_0(dq)\,\Omega_0(dk),
\end{equation}
and therefore by Eqns. \ref{eq:gluon_bubble_1} and \ref{eq:Xi}
\begin{equation}
\Pi^{(g)ab\mu\nu}(\Gamma)=-\frac{3}{32\pi^2}g_s^2\delta^{ab}\int\chi_{\Gamma}(q+k)M^{\mu\nu}(q,k)\,\Omega_0(dq)\,\Omega_0(dk), \nonumber 
\end{equation}
where
\begin{align*}
M^{\mu\nu}(q,k)  &= N^{\mu\nu}(q+k,-k) \\
    &=\eta_{\alpha\beta}\eta_{\rho\sigma}[\eta^{\mu\alpha}(q+2k)^{\rho}+\eta^{\alpha\rho}(q-k)^{\mu}-\eta^{\rho\mu}(2q+k)^{\alpha}] \\
    &  \times[\eta^{\nu\beta}(q+2k)^{\sigma}+\eta^{\beta\sigma}(q-k)^{\nu}-\eta^{\sigma\nu}(2q+k)^{\beta}] \\
    & = \eta_{\alpha\beta}\eta_{\rho\sigma}\eta^{\mu\alpha}\eta^{\nu\beta}(q+2k)^{\rho}(q+2k)^{\sigma} \\
    & +\eta_{\alpha\beta}\eta_{\rho\sigma}\eta^{\mu\alpha}\eta^{\beta\sigma}(q+2k)^{\rho}(q-k)^{\nu} \\
    & -\eta_{\alpha\beta}\eta_{\rho\sigma}\eta^{\mu\alpha}\eta^{\sigma\nu}(q+2k)^{\rho}(2q+k)^{\beta} \\
    & +\eta_{\alpha\beta}\eta_{\rho\sigma}\eta^{\alpha\rho}\eta^{\nu\beta}(q-k)^{\mu}(q+2k)^{\sigma} \\
    & +\eta_{\alpha\beta}\eta_{\rho\sigma}\eta^{\alpha\rho}\eta^{\beta\sigma}(q-k)^{\mu}(q-k)^{\nu} \\
    & -\eta_{\alpha\beta}\eta_{\rho\sigma}\eta^{\alpha\rho}\eta^{\sigma\nu}(q-k)^{\mu}(2q+k)^{\beta} \\
    & -\eta_{\alpha\beta}\eta_{\rho\sigma}\eta^{\rho\mu}\eta^{\nu\beta}(2q+k)^{\alpha}(q+2k)^{\sigma} \\
    & -\eta_{\alpha\beta}\eta_{\rho\sigma}\eta^{\rho\mu}\eta^{\beta\sigma}(2q+k)^{\alpha}(q-k)^{\nu} \\
    & +\eta_{\alpha\beta}\eta_{\rho\sigma}\eta^{\rho\mu}\eta^{\sigma\nu}(2q+k)^{\alpha}(2q+k)^{\beta} \\
    & = (q+2k)^2\eta^{\mu\nu} \\
    &  +(q+2k)^{\mu}(q-k)^{\nu} \\
    &  -(2q+k)^{\mu}(q+2k)^{\nu} \\
    &  +(q-k)^{\mu}(q+2k)^{\nu} \\
    &  +4(q-k)^{\mu}(q-k)^{\nu} \\
    &  -(q-k)^{\mu}(2q+k)^{\nu} \\
    &  -(q+2k)^{\mu}(2q+k)^{\nu} \\
    &  -(2q+k)^{\mu}(q-k)^{\nu} \\
    &  +(2q+k)^2\eta^{\mu\nu}\\
  =&(5q^2+5k^2+8q\cdot k)\eta^{\mu\nu}+q^{\mu}q^{\nu}+2k^{\mu}q^{\nu}-q^{\mu}k^{\nu}-2k^{\mu}k^{\nu}-2q^{\mu}q^{\nu}-k^{\mu}q^{\nu}-4q^{\mu}k^{\nu}\\
&-2k^{\mu}k^{\nu}+q^{\mu}q^{\nu}-k^{\mu}q^{\nu}+2q^{\mu}k^{\nu}-2k^{\mu}k^{\nu}+4q^{\mu}q^{\nu}-4k^{\mu}q^{\nu}-4q^{\mu}k^{\nu}+4k^{\mu}k^{\nu}\\
&-2q^{\mu}q^{\nu}-q^{\mu}k^{\nu}+2k^{\mu}q^{\nu}+k^{\mu}k^{\nu}-2q^{\mu}q^{\nu}-4k^{\mu}q^{\nu}-q^{\mu}k^{\nu}-2k^{\mu}k^{\nu}\\
&-2q^{\mu}q^{\nu}-k^{\mu}q^{\nu}+2q^{\mu}k^{\nu}+k^{\mu}k^{\nu}\\
=&(5q^2+5k^2+8q\cdot k)\eta^{\mu\nu}-2q^{\mu}q^{\nu}-2k^{\mu}k^{\nu}-7q^{\mu}k^{\nu}-7k^{\mu}q^{\nu}.
\end{align*}
It is straightforward to show that $\Pi^{(g)ab\mu\nu}$ is a well defined Borel measure on Minkowski space ${\bf R}^4$.
Write
\begin{equation}\label{eq:Pi_gluon_total}
\Pi^{(g)ab\mu\nu}(\Gamma)=\Pi^{(g)ab}\Pi^{(g)\mu\nu}(\Gamma),
\end{equation}
where
\begin{equation}\nonumber
\Pi^{(g)ab}= -\frac{3}{32\pi^2}g_s^2\delta^{ab},
\end{equation}
and
\begin{equation}\nonumber
\Pi^{(g)\mu\nu}(\Gamma)=\int\chi_{\Gamma}(q+k)M^{\mu\nu}(q,k)\,\Omega_0(dq)\,\Omega_0(dk).
\end{equation}
Using the Lorentz invariance of $\Omega_0$ we have that, for any $\Lambda\in O(1,3)^{\uparrow+}, \Gamma\in{\mathcal B}_0({\bf R}^4)$ 
\begin{align*}
\Pi^{(g)\mu\nu}(\Lambda\Gamma)&= \int\chi_{(\Lambda\Gamma)}(q+k)M^{\mu\nu}(q,k)\,\Omega_0(dq)\,\Omega_0(dk),\\
&=\int\chi_{\Gamma}(\Lambda^{-1}q+\Lambda^{-1}k)M^{\mu\nu}(q,k)\,\Omega_0(dq)\,\Omega_0(dk),\\
&=\int\chi_{\Gamma}(q+k)M^{\mu\nu}(\Lambda q,\Lambda k)\,\Omega_0(dq)\,\Omega_0(dk).
\end{align*} 
But we have
\begin{align*}
M^{\mu\nu}(\Lambda q,\Lambda k)&=\eta^{\mu\nu}(5(\Lambda q)^2+5(\Lambda k)^2+8(\Lambda q)\cdot(\Lambda k))-2(\Lambda q)^{\mu}(\Lambda q)^{\nu}-\\
&2(\Lambda k)^{\mu}(\Lambda k)^{\nu}-7(\Lambda q)^{\mu}(\Lambda k)^{\nu}-7(\Lambda k)^{\mu}(\Lambda q)^{\nu}\\
&=\Lambda^{\mu}{}_{\rho}\Lambda^{\nu}{}_{\sigma}(\eta^{\rho\sigma}(5q^2+5k^2+8q\cdot k)-2q^{\rho}q^{\sigma}-\\
&2k^{\rho}k^{\sigma}-7q^{\rho}k^{\sigma}-7k^{\rho}q^{\sigma})\\
&=\Lambda^{\mu}{}_{\rho}\Lambda^{\nu}{}_{\sigma}M^{\rho\sigma}(q,k).
\end{align*}
Therefore
\[ \Pi^{(g)\mu\nu}(\Lambda\Gamma)=\Lambda^{\mu}{}_{\rho}\Lambda^{\nu}{}_{\sigma}\Pi^{(g)\rho\sigma}(\Gamma),\forall\Lambda\in O(1,3)^{\uparrow+},\Gamma\in{\mathcal B}_0({\bf R}^4), \]
and so $\Pi^{\mu\nu}$ is a Lorentz  covariant measure in the sense defined in Ref. \citen{NPB}.

By a similar argument to that used in  Ref. \citen{Symmetry} the support of $\Pi^{(g)\mu\nu}$ satisfies supp$(\Pi^{(g)\mu\nu})\subset\{q\in{\bf R}^4:q^2\geq0, q^0\geq0\}$. Thus $\Pi^{(g)\mu\nu}$ is causal. 
We will now show, using the spectral calculus, \cite{Springer,Symmetry,NPB,Lamb_shift} that $\Pi^{(g)\mu\nu}$ has a spectral representation
\[ \Pi^{(g)\mu\nu}(\Gamma)=\int_{m^{\prime}=0}^{\infty}\int_{q\in\Gamma}q^2\eta^{\mu\nu}\,\Omega_{m^{\prime}}(dq)\,\sigma_1(dm^{\prime})+\int_{m^{\prime}=0}^{\infty}\int_{q\in\Gamma}q^{\mu}q^{\nu}\,\Omega_{m^{\prime}}(dq)\,\sigma_2(dm^{\prime}), \]
where the spectral measures $\sigma_1,\sigma_2$ are continuous functions. Let $a,b\in{\bf R}, 0<a<b$. We have
\begin{align*}
g^{\mu\nu}(a,b,\epsilon) & =\Pi^{(g)\mu\nu}(\Gamma(a,b,\epsilon)) \\
    & =\int\chi_{\Gamma(a,b,\epsilon)}(q+k)M^{\mu\nu}(q,k)\,\Omega_0(dk)\Omega_0(dq) \\
    & \approx\int\chi_{(a,b)}(|{\vct q}|+|{\vct k}|)\chi_{B(\epsilon,{\vct 0})}({\vct q}+{\vct k})M^{\mu\nu}((|{\vct q}|,{\vct q}),(|{\vct k}|,{\vct k}))\,\frac{d{\vct k}}{|{\vct k}|}\frac{d{\vct q}}{|{\vct q}|} \\
    & =\int\chi_{(a,b)}(|{\vct q}|+|{\vct k}|)\chi_{B(\epsilon,{\vct 0})-{\vct q}}({\vct k})M^{\mu\nu}((|{\vct q}|,{\vct q}),(|{\vct k}|,{\vct k}))\,\frac{d{\vct k}}{|{\vct k}|}\frac{d{\vct q}}{|{\vct q}|} \\
    & \approx\int\chi_{(a,b)}(2|{\vct q}|)M^{\mu\nu}((|{\vct q}|,{\vct q}),(|{\vct q}|,-{\vct q}))\,\frac{d{\vct q}}{|{\vct q}|^2}(\frac{4}{3}\pi\epsilon^3),
\end{align*}
for all $a,b\in{\bf R}$ with $0<a<b$, where
\[ B(\epsilon,{\vct0})=B_{\epsilon}({\vct0})=\{{\vct q}\in{\bf R}^3:|{\vct q}|<\epsilon\}, \] 
is the ball of radius $\epsilon$ centered on the origin ${\vct0}\in{\bf R}^3$. Now 
\begin{align*}
\chi_{(a,b)}(2|{\vct q}|)=1&\Leftrightarrow2|{\vct q}|\in(a,b)\Leftrightarrow|{\vct q}|\in(\frac{a}{2},\frac{b}{2}).
\end{align*}
Hence, using spherical polar coordinates,
\begin{align*}
g_a^{\mu\nu}(b) &=\int_{s=\frac{a}{2}}^{\frac{b}{2}}\int_{\theta=0}^{\pi}\int_{\phi=0}^{2\pi}M^{\mu\nu}((s,s\sin(\theta)\cos(\phi),s\sin(\theta)\sin(\phi),s\cos(\theta)),\\
&(s,-s\sin(\theta)\cos(\phi),-s\sin(\theta)\sin(\phi),-s\cos(\theta)))\sin(\theta)\,d\phi\,d\theta\,ds\,(\frac{4}{3}\pi).
\end{align*}
Thus, using the Leibniz integral rule
\begin{align*}
g_a^{\mu\nu\prime}(b) &=\int_{\theta=0}^{\pi}\int_{\phi=0}^{2\pi}M^{\mu\nu}(\frac{b}{2}(1,\sin(\theta)\cos(\phi),\sin(\theta)\sin(\phi),\cos(\theta)),\\
&\frac{b}{2}(1,-\sin(\theta)\cos(\phi),-\sin(\theta)\sin(\phi),-\cos(\theta)))\sin(\theta)\,d\phi\,d\theta\,(\frac{2}{3}\pi)\\
&=b^2\int_{\theta=0}^{\pi}\int_{\phi=0}^{2\pi}M^{\mu\nu}((1,\sin(\theta)\cos(\phi),\sin(\theta)\sin(\phi),\cos(\theta)),\\
&(1,-\sin(\theta)\cos(\phi),-\sin(\theta)\sin(\phi),-\cos(\theta)))\sin(\theta)\,d\phi\,d\theta\,(\frac{1}{6}\pi),
\end{align*}
where we have used the fact that
\[ M(\lambda q,\lambda k)=\lambda^2M(q,k),\forall\lambda\in{\bf R},q,k\in{\bf R}^4. \]
Therefore, since if 
\[ q=(1,\sin(\theta)\cos(\phi),\sin(\theta)\sin(\phi),\cos(\theta)), \]
and 
\[ k=(1,-\sin(\theta)\cos(\phi),-\sin(\theta)\sin(\phi),-\cos(\theta)), \]
then $q^2=k^2=0$, $q\cdot k=2$, $k=(k^0,{\vct k})=(q^0,-{\vct q})$, we have, for all $i\in\{1,2,3\}$, 
\begin{align*}
M^{ii}(q,k)&=-5q^2-5k^2-8q\cdot k-2(q^{i})^2-2(k^{i})^2-7q^{i}k^{i}-7k^{i}q^{i}\\
&=-16-2(q^{i})^2-2(q^{i})^2+7(q^{i})^2+7(q^{i})^2\\
&=-16+10(q^{i})^2,
\end{align*}
and
\[ M^{00}(q,k)=5q^2+5k^2+8q\cdot k-2(q^0)^2-2(q^0)^2-7(q^0)^2-7(q^0)^2=16-18(q^0)^2=-2. \]
Therefore
\begin{align*}
g_a^{11\prime}(b) &=\frac{1}{6}\pi b^2\int_{\theta=0}^{\pi}\int_{\phi=0}^{2\pi}(-16+10\sin^2(\theta)\cos^2(\phi))\sin(\theta)\,d\phi\,d\theta\\
&=\frac{1}{6}\pi b^2(-16(4\pi)+10(\frac{4}{3})\pi)\\
&=-\frac{76}{9}\pi^2 b^2,\\
g_a^{22\prime}(b)&=\frac{1}{6}\pi b^2\int_{\theta=0}^{\pi}\int_{\phi=0}^{2\pi}(-16+10\sin^2(\theta)\sin^2(\phi))\sin(\theta)\,d\phi\,d\theta\\
&=\frac{1}{6}\pi b^2(-16(4\pi)+10(\frac{4}{3})\pi)\\
&=-\frac{76}{9}\pi^2 b^2,\\
g_a^{33\prime}(b) &=\frac{1}{6}\pi b^2\int_{\theta=0}^{\pi}\int_{\phi=0}^{2\pi}(-16+10\cos^2(\theta))\sin(\theta)\,d\phi\,d\theta\\
&=\frac{1}{6}\pi b^2(2\pi)(-16(2)+10(\frac{2}{3}))\\
&=-\frac{76}{9}\pi^2 b^2.
\end{align*}
Thus 
\[ g_a^{ii\prime}(b)=g_a^{jj\prime}(b),\forall i,j\in\{1,2,3\},a,b\text{ with }b>a>0, \]
and
as it should from the general theory \cite{Lamb_shift} and
\begin{align}
g_a^{ii\prime}(b)=-\frac{76}{9}\pi^2 b^2,\forall i\in\{1,2,3\},a,b\text{ with } b>a>0.\label{eq:g_ii_prime}
\end{align} 
Also
\begin{align}
g_a^{00\prime}(b)=&\frac{1}{6}\pi b^2\int_{\theta=0}^{\pi}\int_{\phi=0}^{2\pi}(-2)\sin(\theta)\,d\phi\,d\theta=-\frac{4}{3}\pi^2b^2,\forall a,b\text{ with }b>a>0.\label{eq:g_a_00_prime}
\end{align}
From the general theory (the spectral theory for Lorentz covariant tensor valued measures \cite{IJMPA,NPB,Lamb_shift}) the measure $\Pi^{(g)\mu\nu}$ is given by
\[ \Pi^{(g)\mu\nu}(\Gamma)=\int_{{m^{\prime}}=0}^{\infty}\int_{q\in{\bf R}^4}\chi_{\Gamma}(q)(\eta^{\mu\nu}q^2\sigma^{(g)}_1(m^{\prime})+q^{\mu}q^{\nu}\sigma^{(g)}_2(m^{\prime}))\,\Omega_{m^{\prime}}(dq)\,dm^{\prime}, \]
where
\begin{align}
\sigma_1^{(g)}(b)&=-\frac{3}{4\pi}\frac{1}{b}g_a^{ii\prime}(b) ,i\in\{1,2,3\},\label{eq:sigma_g_1}\\
\sigma_2^{(g)}(b)&=\frac{3}{4\pi}\frac{1}{b}g_a^{00\prime}(b)-\sigma_1(b),\label{eq:sigma_g_2} 
\end{align}
and is associated with a density $\Pi^{(g)\mu\nu}:\{q\in{\bf R}^4:q^2>0,q^0>0\}\rightarrow{\bf R}$ given by
\begin{align*}
\Pi^{(g)\mu\nu}(q)&=\eta^{\mu\nu}s\sigma_1^{(g)}(s)+q^{\mu}q^{\nu}s^{-1}\sigma_2^{(g)}(s),
\end{align*}
for $q^2>0,q^0>0$ where $s=(q^2)^{\frac{1}{2}}$.
From Eqns.~\ref{eq:g_ii_prime}, ~\ref{eq:g_a_00_prime}, \ref{eq:sigma_g_1} and \ref{eq:sigma_g_2} we have that
\[ \sigma^{(g)}_1(s)=(-\frac{3}{4\pi}\frac{1}{s})(-\frac{76}{9}\pi^2s^2)=\frac{19}{3}\pi s,\forall s>0, \]
and
\begin{align*}
\sigma_2^{(g)}(s)=&\frac{3}{4\pi}\frac{1}{s}g_a^{00\prime}(s)-\sigma_1(s)=(\frac{3}{4\pi}\frac{1}{s})(-\frac{4}{3}\pi^2s^2)-\sigma_1(s)\\
=&-\pi s-\sigma_1(s)=-\frac{22}{3}\pi s,\forall s>0.
\end{align*}
Therefore
\begin{align*}
\Pi^{(g)\mu\nu}(q) =&\frac{19}{3}\pi s^2\eta^{\mu\nu}+q^{\mu}q^{\nu}s^{-1}\sigma_2^{(g)}(s).
 \end{align*}
In the spacelike domain, since both $\sigma_1$ and $\sigma_2$ are odd functions, $\Pi^{(g)\mu\nu}:\{q\in{\bf R}^4:q^2<0\}\rightarrow{\bf R}$ is given by
\begin{align*}
\Pi^{(g)\mu\nu}(q) &=\frac{19}{3}\pi Q^2\eta^{\mu\nu}+q^{\mu}q^{\nu}Q^{-1}\sigma_2^{(g)}(Q),
\end{align*}
where $Q=(-q^2)^{\frac{1}{2}}$.

Hence, from Eq.~\ref{eq:Pi_gluon_total}
\begin{align*}
\Pi^{(g)ab\mu\nu}(q)=&(-\frac{3}{32\pi^2}g_s^2\delta^{ab})(\frac{19}{3}\pi Q^2\eta^{\mu\nu}+Q^{-1}\sigma_2^{(g)}(Q)q^{\mu}q^{\nu})\\
=&-\frac{19}{32\pi}g_s^2\delta^{ab}\eta^{\mu\nu}Q^2-\frac{3}{32\pi^2}g_s^2\delta^{ab}Q^{-1}\sigma_2^{(g)}(Q)q^{\mu}q^{\nu},
\end{align*}
where $Q=(-q^2)^{\frac{1}{2}}$.

\section{The four-point gluon bubble\label{section:4-point_gluon_bubble}}

The four-point gluon bubble is associated with a function $q\mapsto \Pi^{(4)ab\mu\nu}(q)$ whose value is a constant tensor times the constant ``function"
\[ f(q)=c\text{ where }c=\int\frac{1}{k^2+i\epsilon}\,dk, \]
(see p. 519 of Ref. \citen{Schwartz}). We will argue (somewhat formally, since the integral defining $c$ is divergent) that $c=0$ as follows. Assume that $c$ is finite. Then we have 
\[ c=\int(k^2+i\epsilon)^{-1}\,dk. \]
Make a change of variables $l=\lambda k$ where $\lambda>0$. Then 
\[ k=\lambda^{-1}l,dk=\lambda^{-4}dl,k^2=\lambda^{-2}l^2. \]
Therefore
\[ c=\int(\lambda^{-2}l^2+i\epsilon)^{-1}\,(\lambda^{-4}dl)=\lambda^{-2}\int(l^2+i\epsilon)^{-1}\,dl=\lambda^{-2}c. \]
Since this is true for all $\lambda>0$ we must have that $c=0$.

Thus the four-point gluon bubble vanishes.

\section{The Momentum Space Spectral QCD Running Coupling}

The method we use is exact and fully relativistic (not an NR approximation). It is not based on approximate solutions to the Lippmann-Schwinger equation (e.g. the Born approximations) and it does not involve the Schwinger-Dyson equation. It applies over the whole range of possible input and output momenta and polarizations. Thus it applies over the whole range of energies, from infrared to ultraviolet. 

To compute the running coupling at one-loop level we compare the process described by the tree level Feynman amplitude ${\mathcal M}^{\text{(tree)}}$ with the process described by the Feynman amplitude ${\mathcal M}={\mathcal M}^{\text{(tree)}}+{\mathcal M}^{\text{(vp)}}$, where ${\mathcal M}^{\text{(vp)}}$ is the one-loop spectral vacuum polarization Feynman amplitude, for quark $u\overline{d}$ scattering.

\subsection{Computation of the Feynman amplitudes ${\mathcal M}^{(\text{tree})}$ and ${\mathcal M}={\mathcal M}^{(\text{tree})}+{\mathcal M}^{(\text{vp})}$}

The tree level process is the process associated with the Feynman diagram given in Fig.~\ref{fig:tree_level}  and, using the Feynman rules, we have that the tree level Feynman amplitude ${\mathcal M}^{(\text{tree})}$ is given by
\begin{align}\label{eq:tree_level_def}
i{\mathcal M}^{\text{(tree)}}_{i_1^{\prime}i_2^{\prime}\alpha_1^{\prime}\alpha_2^{\prime}i_1i_2\alpha_1\alpha_2}(p_1^{\prime},p_2^{\prime},p_1,p_2)&=\overline{u}(p_1^{\prime},\alpha_1^{\prime})ig_sT^{a}_{i_1^{\prime}i_1}\gamma^{\mu}u(p_1,\alpha_1)\\
&(-i\frac{\delta_{ab}\eta_{\mu\nu}}{q^2+i\epsilon})\nonumber\\
&\overline{v}(p_2,\alpha_2)ig_sT^{b}_{i_2i_2^{\prime}}\gamma^{\nu}v(p_2^{\prime},\alpha_2^{\prime})\nonumber,
\end{align}
where $u$ and $v$ are the Dirac spinors defined by (with the Dirac equation considered with respect to the chiral (or Weyl) representation of the Dirac gamma matrices)
\begin{align*}
&u(p,\alpha)=\left(\begin{array}{c}
(p\cdot\sigma)^{\frac{1}{2}}e_{\alpha}\\
(p\cdot\overline{\sigma})^{\frac{1}{2}}e_{\alpha}
\end{array}\right),\\
&v(p,\alpha)=\left(\begin{array}{c}
(p\cdot\sigma)^{\frac{1}{2}}e_{\alpha}\\
-(p\cdot\overline{\sigma})^{\frac{1}{2}}e_{\alpha}
\end{array}\right),
\end{align*}
for $p\in {\bf R}^4,\alpha\in\{1,2\}$ in which $\sigma=(\sigma_0,\sigma_1,\sigma_2,\sigma_3),\overline{\sigma}=(\sigma_0,-\sigma_1,-\sigma_2,-\sigma_3)$ where $\sigma_0=1$ is the $2\times2$ unit matrix and $\{\sigma_i\}_{i=1}^3$ are the Pauli sigma matrices, $\{e_{\alpha}\}_{\alpha=1,2}$ is the standard basis for ${\bf C}^2$ and $q=p_1^{\prime}-p_1$ is the momentum transfer.

We have the following well known lemma.
\begin{lemma} If the incoming and outgoing up quark momenta $p_1$ and $p_1^{\prime}$ are on shell and ${\vct p}_1\neq{\vct0}$ then the momentum transfer $q=p_1^{\prime}-p_1$ is spacelike with $q^2<0$.
\end{lemma}
From Eq.~\ref{eq:tree_level_def} we have that
\begin{align}
&{\mathcal M}^{\text{(tree)}}_{i_1^{\prime}i_2^{\prime}\alpha_1^{\prime}\alpha_2^{\prime}i_1i_2\alpha_1\alpha_2}(p_1^{\prime},p_2^{\prime},p_1,p_2)\nonumber\\
&=-\frac{g_s^2}{Q^2}\delta_{ab}\eta_{\mu\nu}T^{a}_{i_1^{\prime}i_1}T^{b}_{i_2i_2^{\prime}}\overline{u}(p_1^{\prime},\alpha_1^{\prime})\gamma^{\mu}u(p_1,\alpha_1)\overline{v}(p_2,\alpha_2)\gamma^{\nu}v(p_2^{\prime},\alpha_2^{\prime})\nonumber\\
&=-\frac{g_s^2}{Q^2}\delta_{ab}\eta_{\mu\nu}T^{a}_{i_1^{\prime}i_1}T^{b}_{i_2i_2^{\prime}}{\mathcal M}_{0,i_1^{\prime}i_2^{\prime}\alpha_1^{\prime}\alpha_2^{\prime}i_1i_2\alpha_1\alpha_2}^{\mu\nu}(p_1^{\prime},p_2^{\prime},p_1,p_2),\label{eq:tree_amplitude}
\end{align}
where $Q=(-q^2)^{\frac{1}{2}}$ and
\[ {\mathcal M}_{0,i_1^{\prime}i_2^{\prime}\alpha_1^{\prime}\alpha_2^{\prime}i_1i_2\alpha_1\alpha_2}^{\mu\nu}(p_1^{\prime},p_2^{\prime},p_1,p_2)=\overline{u}(p_1^{\prime},\alpha_1^{\prime})\gamma^{\mu}u(p_1,\alpha_1)\overline{v}(p_2,\alpha_2)\gamma^{\nu}v(p_2^{\prime},\alpha_2^{\prime}). \]
The vacuum polarization insertion is as shown in the Feynman diagram of Fig.~\ref{fig:vacuum_polarization} (this diagram stands for two diagrams, one with the quark bubble and the other with the gluon bubble). The Feynman amplitude associated with this diagram, ${\mathcal M}^{(\text{vp})}$, is given by \begin{align*}
i{\mathcal M}^{\text{(vp)}}_{i_1^{\prime}i_2^{\prime}\alpha_1^{\prime}\alpha_2^{\prime}i_1i_2\alpha_1\alpha_2}(p_1^{\prime},p_2^{\prime},p_1,p_2)=&\overline{u}(p_1^{\prime},\alpha_1^{\prime})ig_sT^{a}_{i_1^{\prime}i_1}\gamma^{\mu}u(p_1,\alpha_1)\\
&iD_{\text{QCD},ac\mu\rho}(q)i\Pi^{\text{(vp)},cd\rho\sigma}(q)iD_{\text{QCD},db\sigma\nu}(q)\\
&\overline{v}(p_2,\alpha_2)ig_sT^{b}_{i_2i_2^{\prime}}\gamma^{\nu}v(p_2^{\prime},\alpha_2^{\prime}),
\end{align*}
where $D_{\text{QCD}}$ is the gluon propagator given by
\[ D_{\text{QCD},ab\mu\nu}(q)=-\frac{\delta_{ab}\eta_{\mu\nu}}{q^2+i\epsilon}, \]
and $\Pi^{\text{(vp)}}$ is the total vacuum polarization tensor. Therefore
\begin{align*}
{\mathcal M}^{\text{(vp)}}_{i_1^{\prime}i_2^{\prime}\alpha_1^{\prime}\alpha_2^{\prime}i_1i_2\alpha_1\alpha_2}(p_1^{\prime},p_2^{\prime},p_1,p_2)=&\frac{g_s^2}{Q^4}T^{a}_{i_1^{\prime}i_1}T^{b}_{i_2i_2^{\prime}}\eta_{\mu\rho}\eta_{\sigma\nu}\delta_{ac}\delta_{db}\Pi^{(vp)cd\rho\sigma}(q)\\
&{\mathcal M}_{0,i_1^{\prime}i_2^{\prime}\alpha_1^{\prime}\alpha_2^{\prime}i_1i_2,\alpha_1\alpha_2}^{\mu\nu}(p_1^{\prime},p_2^{\prime},p_1,p_2).
 \end{align*}
Now 
\[ \Pi^{(vp)}=\Pi^{(q,\text{tot})}+\Pi^{(g)}+\Pi^{(4)}, \]
From Section~\ref{section:4-point_gluon_bubble}, $\Pi^{(4)}=0$. Also from Sections \ref{section:quark_bubble} and \ref{section:gluon_bubble} we have
\[ \Pi^{(q,\text{tot})ab\mu\nu}(q)=-\frac{1}{3\pi}g_s^2\delta^{ab}\sum_{k=1}^6f(m_k,Q)(Q^2\eta^{\mu\nu}+q^{\mu}q^{\nu}), \]
 and
\begin{align*}
\Pi^{(g)ab\mu\nu}(q)=&-\frac{19}{32\pi}g_s^2\delta^{ab}\eta^{\mu\nu}Q^2-\frac{3}{32\pi^2}g_s^2\delta^{ab}q^{\mu}q^{\nu}Q^{-1}\sigma_2^{(g)}(Q),
\end{align*}
where
\[ f(m,Q)=m^3Q^{-3}Z(m,Q)(3+2Z(m,Q)^2), \]
and $Q=(-q^2)^{\frac{1}{2}}$.

By a well known conservation property (see p. 480 of Ref. \citen{Weinberg}), we can, in the presence of the factor ${\mathcal M}_0^{\mu\nu}$, drop the terms involving $q_{\mu}q_{\nu}=\eta_{\mu\rho}q^{\rho}q^{\sigma}\eta_{\sigma\nu}$. Therefore
\begin{align*}
&{\mathcal M}^{\text{(vp)}}_{i_1^{\prime}i_2^{\prime}\alpha_1^{\prime}\alpha_2^{\prime}i_1i_2\alpha_1\alpha_2}(p_1^{\prime},p_2^{\prime},p_1,p_2)\\
=&\frac{g_s^2}{Q^2}T^{a}_{i_1^{\prime}i_1}T^{b}_{i_2i_2^{\prime}}\eta_{\mu\rho}\eta_{\sigma\nu}\delta_{ac}\delta_{db}\\
&(-\frac{1}{3\pi}g_s^2\delta^{cd}\eta^{\rho\sigma}\sum_{k=1}^6f(m_k,Q)-\frac{19}{32\pi}g_s^2\delta^{cd}\eta^{\rho\sigma})\\
&\times{\mathcal M}_{0,i_1^{\prime}i_2^{\prime}\alpha_1^{\prime}\alpha_2^{\prime}i_1i_2\alpha_1\alpha_2}^{\mu\nu}(p_1^{\prime},p_2^{\prime},p_1,p_2)\\
=&-\frac{g_s^2}{Q^2}T^{a}_{i_1^{\prime}i_1}T^{a}_{i_2i_2^{\prime}}{\mathcal M}_{0,i_1^{\prime}i_2^{\prime}\alpha_1^{\prime}\alpha_2^{\prime}i_1i_2\alpha_1\alpha_2}(p_1^{\prime},p_2^{\prime},p_1,p_2)\pi^{(s)}(Q),
\end{align*}
where
\[ \pi^{(s)}(Q)=g_s^2(\frac{19}{32\pi}+\frac{1}{3\pi}\sum_{k=1}^6f(m_k,Q)), \]
and
\[ {\mathcal M}_0=\eta_{\mu\nu}{\mathcal M}_0^{\mu\nu}. \]
The total Feynman amplitude ${\mathcal M}={\mathcal M}^{(\text{tree})}+{\mathcal M}^{(\text{vp})}$ is given by
\begin{align*}
&{\mathcal M}_{i_1^{\prime}i_2^{\prime}\alpha_1^{\prime}\alpha_2^{\prime}i_1i_2\alpha_1\alpha_2}(p_1^{\prime},p_2^{\prime},p_1,p_2)\\
=&-\frac{g_s^2}{Q^2}T^{a}_{i_1^{\prime}i_1}T^{a}_{i_2i_2^{\prime}}{\mathcal M}_{0,i_1^{\prime}i_2^{\prime}\alpha_1^{\prime}\alpha_2^{\prime}i_1i_2\alpha_1\alpha_2}(p_1^{\prime},p_2^{\prime},p_1,p_2)(1+\pi^{(s)}(Q)).
\end{align*}

\subsection{Computation of the quantity $\Gamma=\Gamma(q)$ such that ${\mathcal M}=\Gamma{\mathcal M}^{(\text{tree})}$}

Consider the quantity $\Gamma:\{q\in{\bf R}^4:q^2<0\}\rightarrow{\bf C}$, a function of the momentum transfer $q$ which, if it exists, specifies over all  values of the input and output momenta and polarizations, a relationship of proportionality between the total Feynman amplitude ${\mathcal M}$ and the tree level amplitude ${\mathcal M}^{(\text{tree})}$. $\Gamma$ is given by the following equation (if the solution of the equation exists)
\[ {\mathcal M}=\Gamma(q){\mathcal M}^{(\text{tree})}. \]
This is equivalent to the following
\begin{align}\label{eq:Gamma_def_0}
&-\frac{g_s^2}{Q^2}T^{a}_{i_1^{\prime}i_1}T^{a}_{i_2i_2^{\prime}}{\mathcal M}_{0,i_1^{\prime}i_2^{\prime}\alpha_1^{\prime}\alpha_2^{\prime}i_1i_2\alpha_1\alpha_2}(p_1^{\prime},p_2^{\prime},p_1,p_2)(1+\pi^{(s)}(Q))\nonumber\\
=&-\Gamma(q)\frac{g_s^2}{Q^2}T^{a}_{i_1^{\prime}i_1}T^{a}_{i_2i_2^{\prime}}{\mathcal M}_{0,i_1^{\prime}i_2^{\prime}\alpha_1^{\prime}\alpha_2^{\prime}i_1i_2\alpha_1\alpha_2}(p_1^{\prime},p_2^{\prime},p_1,p_2),
\end{align}
for all sets of momenta and polarizations.
 
Suppose that $\Gamma:\{q\in{\bf R}^4:q^2<0\}\rightarrow{\bf C}$ were a function such that Eq.~\ref{eq:Gamma_def_0} is true for all sets of momenta and polarizations. Then we can cancel 
\[ X=-\frac{g_s^2}{Q^2}T^{a}_{i_1^{\prime}i_1}T^{a}_{i_2i_2^{\prime}}{\mathcal M}_{0,i_1^{\prime}i_2^{\prime}\alpha_1^{\prime}\alpha_2^{\prime}i_1i_2\alpha_1\alpha_2}(p_1^{\prime},p_2^{\prime},p_1,p_2), \]
from both sides of the equation resulting in the equation
\begin{equation}\label{eq:cancelled_Gamma_equation}
\Gamma(q)=1+\pi^{(s)}(Q),
\end{equation}
where $Q=(-q^2)^{\frac{1}{2}}$. Conversely suppose that $\Gamma:\{q\in{\bf R}^4:q^2<0\}\rightarrow{\bf C}$ is given by Eq.~\ref{eq:cancelled_Gamma_equation}. Then, for any set of values
\[ (p_1^{\prime},p_2^{\prime},p_1,p_2,i_1^{\prime},i_2^{\prime},\alpha_1^{\prime},\alpha_2^{\prime},i_1,i_2,\alpha_1,\alpha_2), \]
of momenta and polarizations we can multiply both sides of Eq.~\ref{eq:cancelled_Gamma_equation} by $X$ to obtain Eq.~\ref{eq:Gamma_def_0}. Thus Eq.~\ref{eq:Gamma_def_0} has the unique solution
\begin{equation}\label{eq:untransformed_Gamma}
\Gamma(q)=1+\pi^{(s)}(Q)=1+g_s^2(\frac{19}{32\pi}+\frac{1}{3\pi}\sum_{k=1}^6f(m_k,Q)),
\end{equation}
for $Q=(-q^2)^{\frac{1}{2}}$ and we see that $\Gamma$ exists and is real valued. 
We can write, without fear of confusion,
\[ \Gamma(Q)=\Gamma(g_s,Q)=1+g_s^2(\frac{19}{32\pi}+\frac{1}{3\pi}\sum_{k=1}^6f(m_k,Q)), \]
for $Q>0$.

The quantity $\Gamma(q)$ serves, for any given value $q$ of the momentum transfer, as a scaling factor reflecting the one-loop corrections within the framework of the spectral regularization method.

\subsection{Computation of the momentum space QCD coupling $q\mapsto g_s(q)$}

From Eq.~\ref{eq:tree_amplitude} we may write that the momentum space QCD coupling $q\mapsto g_s(q)$ as a function of momentum transfer $q$ is given by
\begin{align}
&{\mathcal M}_{i_1^{\prime}i_2^{\prime}\alpha_1^{\prime}\alpha_2^{\prime}i_1i_2\alpha_1\alpha_2}(p_1^{\prime},p_2^{\prime},p_1,p_2)\nonumber\\
&=-\frac{g_s(q)^2}{Q^2}T^{a}_{i_1^{\prime}i_1}T^{a}_{i_2i_2^{\prime}}{\mathcal M}_{0,i_1^{\prime}i_2^{\prime}\alpha_1^{\prime}\alpha_2^{\prime}i_1i_2\alpha_1\alpha_2}(p_1^{\prime},p_2^{\prime},p_1,p_2),\label{eq:M_1}
\end{align}
that is if, indeed, there exists a function $q\mapsto g_s(q)$ of momentum transfer such that the total Feynman amplitude ${\mathcal M}$ can be written in this form. But we have just shown that
\begin{align}
&{\mathcal M}_{i_1^{\prime}i_2^{\prime}\alpha_1^{\prime}\alpha_2^{\prime}i_1i_2\alpha_1\alpha_2}(p_1^{\prime},p_2^{\prime},p_1,p_2)\nonumber\\
=&\Gamma(q){\mathcal M}^{\text{(tree)}}_{i_1^{\prime}i_2^{\prime}\alpha_1^{\prime}\alpha_2^{\prime}i_1i_2\alpha_1\alpha_2}(p_1^{\prime},p_2^{\prime},p_1,p_2)\nonumber\\
=&-\Gamma(q)\frac{g_s^2}{Q^2}T^{a}_{i_1^{\prime}i_1}T^{a}_{i_2i_2^{\prime}}{\mathcal M}_{0,i_1^{\prime}i_2^{\prime}\alpha_1^{\prime}\alpha_2^{\prime}i_1i_2\alpha_1\alpha_2}(p_1^{\prime},p_2^{\prime},p_1,p_2)\label{eq:M_2}.
\end{align}
Therefore we may equate the right hand sides of Eqns.~\ref{eq:M_1} and \ref{eq:M_2} and make cancellations to show that the (one-loop) momentum space QCD coupling $q\mapsto g_s(q)$ is given by
\[ g_s(q)^2=g_s^2\Gamma(q)\text{ and }g_s(Q)^2=g_s^2\Gamma(Q). \]
We denote, from now on, the constant $g_s$ occurring in the Feynman amplitude expressions, by $g_b$ and call it the bare coupling constant. In our work $g_b$ is finite. $\Gamma$ depends (implicitly) on $g_b$ (see Eq.~\ref{eq:untransformed_Gamma}).
$\alpha_s(Q)=\frac{g_s(Q)^2}{4\pi}=\frac{g_b^2}{4\pi}\Gamma(Q)=\alpha_b\Gamma(Q)$ is the momentum space running QCD coupling constant, where $\alpha_b=\frac{g_b^2}{4\pi}$ is the bare QCD coupling constant. 

\begin{figure} 
\centering
\includegraphics[width=14cm]{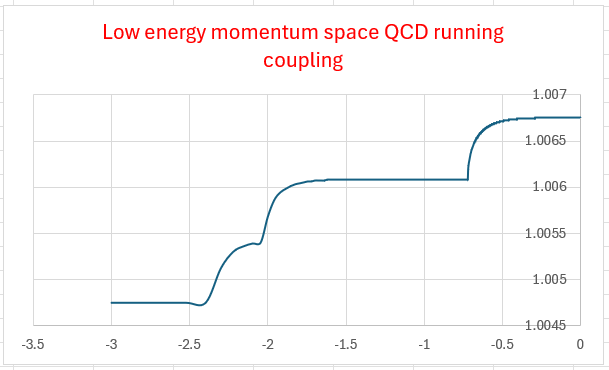}
\caption{Untransformed low energy momentum spectral QCD running coupling \, vs. log of energy in GeV\label{fig:low_en_ms_rc}}
\end{figure}

\begin{figure} 
\centering
\includegraphics[width=14cm]{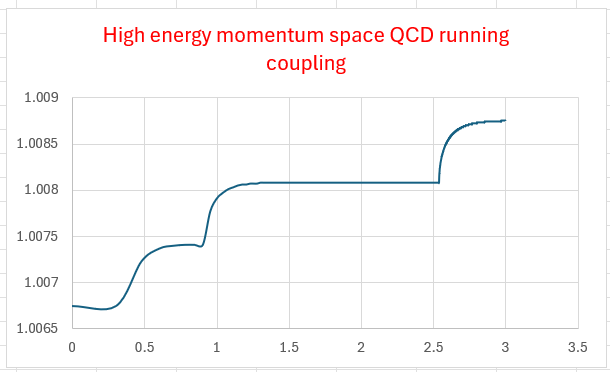}
\caption{Untransformed high energy momentum spectral QCD running coupling vs. log of energy in GeV\label{fig:high_en_ms_rc}}
\end{figure}

The graph of the low energy momentum spectral QCD running coupling (with $\alpha_b=0.002$) is shown in Fig.~\ref{fig:low_en_ms_rc}. The graph of the high energy momentum spectral QCD running coupling (with $\alpha_b=0.002$) is shown in Fig. \ref{fig:high_en_ms_rc}.

$\Gamma$ depends (implicitly) on $\alpha_b$. We will show later how $\alpha_b$ can be determined in a natural fashion and that the value that we will obtain is of the order of 0.002. Therefore, for purposes of illustration we, for the display of the momentum space form of $\Gamma$, set the value of $\alpha_b$ to be 0.002.

We see that the quantity $\Gamma(Q)$ is the scaling factor between the bare strong coupling $\alpha_b$ and the untransformed momentum running coupling $\alpha_s(Q)$.

\section{The Position Space Spectral QCD Running Coupling}

The function $\Gamma:\{q\in{\bf R}^4:q^2<0\}\rightarrow(0,\infty)$ defined by Eq.~\ref{eq:untransformed_Gamma} gives (after multiplication by $\alpha_b$) the untransformed momentum space QCD coupling. The position space QCD coupling is given, formally, by taking the inverse Fourier transform of $q\mapsto\Gamma(q)$ according to
\[ \Gamma(x)=(2\pi)^{-4}\int\Gamma(q)e^{iq\cdot x}\,dq\text{ for } x\in{\bf R}^4. \]
The function $q\mapsto\Gamma(q)$ is Lorentz invariant. Therefore its inverse Fourier transform is formally rotationally invariant on the spacelike hyperplane $x^0=0$. Hence the position space form of $\Gamma$ as a function of distance $r>0$ is given, formally, by
\begin{equation}\label{eq:formal_position_space_rc}
\Gamma(r)=(2\pi)^{-4}\int\Gamma(q)e^{iq\cdot(0,0,0,r)}\,dq.
\end{equation}
$q\mapsto\Gamma(q)$ is not an $L^1$ function or an $L^2$ function and it is straightforward to show that the integral given by  Eq.~\ref{eq:formal_position_space_rc} is not convergent.

Therefore we introduce a momentum-energy cutoff $\Lambda>0$ and write, using spherical polar coordinates for ${\vct q}$, 
\[ q=(q^0,{\vct q})=(E,{\vct q})=(E,\rho\sin(\theta)\cos(\phi),\rho\sin(\theta)\sin(\phi),\rho\cos(\theta)), \] 
\begin{align*}
\Gamma_{\Lambda}(r)=&(2\pi)^{-4}\int_{S_{\Lambda}}\Gamma(q)e^{iq\cdot(0,0,0,r)}\,dq\\
=&(2)(2\pi)^{-4}(2\pi)\int_{\rho=0}^{\Lambda}\int_{E=0}^{\rho}\int_{\theta=0}^{\pi}\Gamma((\rho^2-E^2)^{\frac{1}{2}})e^{-i\rho\cos(\theta)r}\rho^2\sin(\theta)\,d\theta\,dE\,d\rho\\
=&\frac{1}{4\pi^3}\int_{\rho=0}^{\Lambda}\int_{E=0}^{\rho}\Gamma((\rho^2-E^2)^{\frac{1}{2}})\left.\frac{1}{i\rho r}e^{i\rho ru}\right|_{u=-1}^1\rho^2\,dE\,d\rho\\
=&\frac{1}{2\pi^3}\frac{1}{r}\int_{\rho=0}^{\Lambda}\int_{E=0}^{\rho}\Gamma((\rho^2-E^2)^{\frac{1}{2}})\rho\sin(\rho r)\,dE\,d\rho,
\end{align*}
where $S_{\Lambda}=\{q\in{\bf R}^4:|{\vct q}|<\Lambda,q^0<|{\vct q}|\}$. The notation $\Gamma(q)$ in the first line of the above sequence of equations means $\Gamma$ evaluated at a point $q\in{\bf R}^4$ in momentum space. This function is Lorentz invariant. Therefore we can write $\Gamma(q)=\Gamma(Q)$ where $Q=(-q^2)^{\frac{1}{2}}$. The $\Gamma$ written in the subsequent lines of of the above sequence of equations is being used in that role. In this case  $Q=(-q^2)^{\frac{1}{2}}=(\rho^2-E^2)^{\frac{1}{2}}$.

Therefore the (cut off) function $\tau\mapsto\Gamma_{\Lambda}(\tau)$ as a function of (collision) energy $\tau>0$ is given by
\[ \Gamma_{\Lambda}(\tau)=\frac{1}{2\pi^3}\tau\int_{\rho=0}^{\Lambda}\int_{E=0}^{\rho}\Gamma((\rho^2-E^2)^{\frac{1}{2}})\rho\sin(\rho/\tau)\,dE\,d\rho. \] 
Letting $\xi=(\rho^2-E^2)^{\frac{1}{2}}$ we have $\xi^2=\rho^2-E^2$ so, for any fixed $\rho$, $EdE=-\xi d\xi$ from which it follows that $dE=-E^{-1}\xi\,d\xi=-\xi(\rho^2-\xi^2)^{-\frac{1}{2}}\,d\xi$ and hence
\[ \Gamma_{\Lambda}(\tau)=\frac{1}{2\pi^3}\tau\int_{\rho=0}^{\Lambda}\left(\int_{\xi=0}^{\rho}\xi(\rho^2-\xi^2)^{-\frac{1}{2}}\Gamma(\xi)\,d\xi\right)\rho\sin(\rho/\tau)\,d\rho. \] 
(No confusion need arise through using the same symbol for $\Gamma$ in all its guises, i.e. a momentum space function $q\mapsto\Gamma(q)$, a position space function $x\mapsto\Gamma(x)$, a function of distance $r$ and a function of energy $\tau$.)

\begin{figure} 
\centering
\includegraphics[width=14cm]{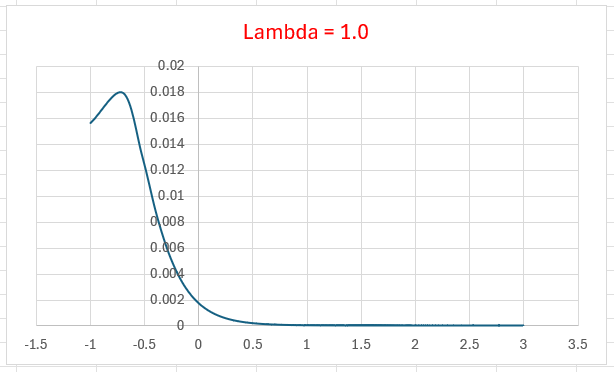}
\caption{Shifted $\Gamma_{\Lambda}$ vs. log of energy in GeV for $\Lambda=1.0$ GeV\label{fig:Lambda_1}}
\end{figure}

\begin{figure} 
\centering
\includegraphics[width=14cm]{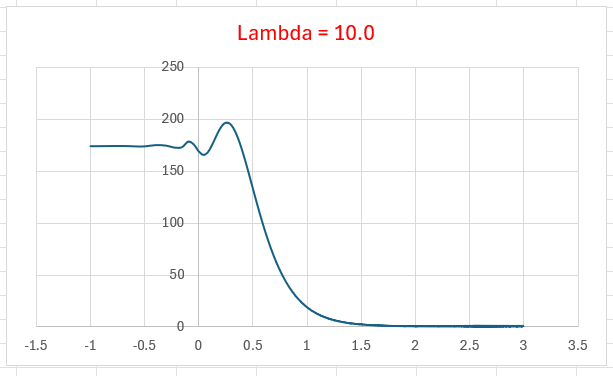}
\caption{Shifted $\Gamma_{\Lambda}$ vs. log of energy in GeV for $\Lambda=10.0$ GeV\label{fig:Lambda_10}}
\end{figure}

\begin{figure} 
\centering
\includegraphics[width=14cm]{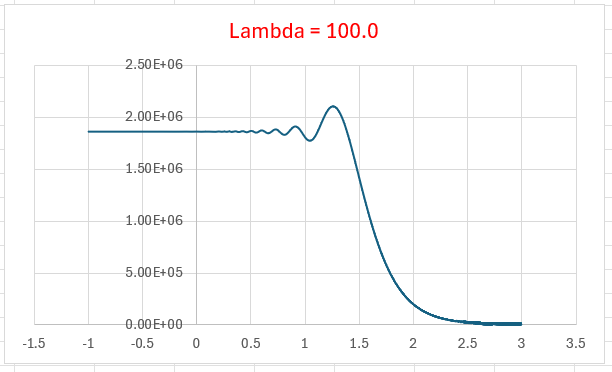}
\caption{Shifted $\Gamma_{\Lambda}$ vs. log of energy in GeV for $\Lambda=100.0$ GeV \label{fig:Lambda_100}}
\end{figure}

Graphs of this function for various values of the cutoff energy $\Lambda$ (and $\alpha_b=1$) are shown in Figs.~\ref{fig:Lambda_1}, \ref{fig:Lambda_10} and \ref{fig:Lambda_100}. We have translated each of these graphs along the y-axis (i.e. the $\alpha_s$-axis) so that their limiting value is zero as $\tau\rightarrow\infty$ (the explanation and justification for this will be given in the next section).

As $\Lambda$ is increased the running coupling graph shifts to the right and the $\alpha_s$ values are scaled up while if $\Lambda$ is decreased the graph moves to the left and the $\alpha_s$ values are scaled down. In all cases the running coupling manifests asymptotic freedom and, while not having a Landau pole, has what might be described as a ``Landau peak" together with subsidiary subpeaks.

\section{Comparison of the Spectral QCD Running Coupling with CERN Data}

The QCD coupling is not an observable and, when its computation involves renormalization, it is renormalization scheme dependent. The renormalization scheme affects the unrenormalized regularized vacuum polarization function by the addition of a constant. For example, with Pauli-Villars regularization the unrenormalized quark bubble vacuum polarization function is given (up to a gluon color factor) by (see p. 323 of Ref. \citen{Itzykson}).
\begin{align*}
\pi(k^2,m,\Lambda)=&-\frac{\alpha}{3\pi}\left\{-\log(\frac{\Lambda^2}{m^2})+\frac{1}{3}+2\left(1+\frac{2m^2}{k^2}\right)\left[\left(\frac{4m^2}{k^2}-1\right)^{\frac{1}{2}}\right.\right.\\
&\left.\left.\times\text{arccot}\left(\frac{4m^2}{k^2}-1\right)^{\frac{1}{2}}-1\right]\right\},
\end{align*}
where $\Lambda$ is the fictitious mass parameter and therefore the counterterm is $\frac{\alpha}{3\pi}\log(\frac{\Lambda^2}{m^2})$. For dimensional regularization the unrenormalized quark bubble vacuum polarization function is given by (see p. 309 of Ref. \citen{Schwartz})
\[ \pi(p^2)=-\frac{g_s^2}{2\pi^2}\int_0^1dx\,x(1-x)\left[\frac{2}{\epsilon}+\log\left(\frac{4\pi e^{-\gamma_E}\mu^2}{m^2-p^2x(1-x)}\right)\right], \]
where $\epsilon$ is the fictitious deviation of the dimension of space-time from 4 and $\mu$ is the subtraction point. For the minimal substraction (MS) renormalization scheme the counterterm involves the factor $\epsilon^{-1}$ while for the modified minimal substraction ($\overline{\text{MS}}$) renormalization scheme the counterterm involves the factor $2\epsilon^{-1}+\log(4\pi e^{-\gamma_E}\mu^2)$. In fact, for the so called ${\mathcal R}_{\delta}$ renormalization scheme \cite{Mojaza} an arbitrary value $\delta$ is added to $\epsilon^{-1}$. In any case the effect of a change in renormalization scheme is the addition of a constant to the counterterm. 

Spectral regularization is finite at all stages and no renormalization is required, there are no counterterms. 

To compare our predictions with data which has been obtained from experimental data using calculations based on the $\overline{\text{MS}}$ renormalization scheme, we need to add a suitable constant to our running coupling function. If all the calculations of the running coupling constant at any given energy based on experiments were done using covariant spectral regularization then no such constant would be required.

Therefore, for the purpose of comparing our spectral running coupling with the $\overline{\text{MS}}$ running coupling or running coupling data interpreted through $\overline{\text{MS}}$ we can write that, with respect to an energy cutoff $\Lambda$, the spectral running coupling $\alpha_{\Lambda}(\tau)$ at energy $\tau$ is
\[ \alpha_{\Lambda}(\tau)=\alpha_b\Gamma_{\Lambda}(\tau)+c, \]
for some $c\in{\bf R}$ and $\alpha_b>0$, where
\[ \Gamma_{\Lambda}(\tau)=\frac{1}{2\pi^3}\tau\int_{\rho=0}^{\Lambda}\left(\int_{\xi=0}^{\rho}\xi(\rho^2-\xi^2)^{-\frac{1}{2}}\Gamma(\xi)\,d\xi\right)\rho\sin(\rho/\tau)\,d\rho. \] 
Since $\Gamma$ depends (implicitly) on $\alpha_b$ we should write $\Gamma_{\Lambda}=\Gamma_{\Lambda}(\alpha_b,\tau)$.

To find the values of $\alpha_b$ and $c$ which give the best (in a certain sense) correspondence between $\alpha_{\Lambda}$  and a collection $\{(\tau_i,\alpha_{i}):i=1,\ldots,n\}$ of running coupling data, such as the CERN data, we note that any function of the form 
\begin{equation}\label{eq:function_form}
\alpha_{\Lambda}(\alpha_b,\tau)=\alpha_n+\frac{\alpha_1-\alpha_n}{\Gamma_{\Lambda}(\alpha_b,\tau_1)-\Gamma_{\Lambda}(\alpha_b,\tau_n)}(\Gamma_{\Lambda}(\alpha_b,\tau)-\Gamma_{\Lambda}(\alpha_b,\tau_n)),
\end{equation}
passes through the points $(\tau_1,\alpha_1)$ and $(\tau_n,\alpha_n)$.
Thus we have scaled $\Gamma_{\Lambda}$ and translated it along the y-axis (i.e. the $\alpha_s$-axis) so that the graph of the function $\alpha_{\Lambda}$ passes through the points $(\tau_1,\alpha_1)$ and $(\tau_n,\alpha_n)$. Eq.~\ref{eq:function_form} is invariant under the transformation
\[ \Gamma_{\Lambda}\mapsto-\Gamma_{\Lambda}. \]
(This is a specific example of the fact that the physics of a process is determined by $|{\mathcal M}|$ where ${\mathcal M}$ is its Feynman amplitude.) Thus we may take the spectral running coupling  at energy $\tau$ to be
\begin{equation}\label{eq:function_form_1}
\alpha_{\Lambda}(\alpha_b,\tau)=-\alpha_b\Gamma_{\Lambda}(\alpha_b,\tau)+c,
\end{equation}
for some $c\in{\bf R},\alpha_b>0$ (we make this transformation because the running coupling $\alpha_s(Q)$ generally decreases as $Q$ increases and we want $\alpha_b$ to be positive). 

Therefore, from Eqns.~\ref{eq:function_form} and \ref{eq:function_form_1} we seek $\alpha_b,c\in{\bf R}$, with $\alpha_b>0$ such that
\begin{equation}\label{eq:alpha_b_computation}
\alpha_b=-(\alpha_1-\alpha_n)(\Gamma_{\Lambda}(\alpha_b,\tau_1)-\Gamma_{\Lambda}(\alpha_b,\tau_n))^{-1}, 
\end{equation}
and
\begin{equation}
c=\alpha_n+\alpha_b\Gamma_{\Lambda}(\alpha_b,\tau_n).\label{eq:c_computation} 
\end{equation}
Analytical solution of this problem would require solution of a complicated integral equation, that is Eq.~\ref{eq:alpha_b_computation}, to determine $\alpha_b$, and then the plugging in of the solution to that equation into Eq.~\ref{eq:c_computation} to determine $c$. Fortunately the problem given by Eq.~\ref{eq:alpha_b_computation} can be solved computationally using the following (fixed point computation) algorithm.
\newtheorem{algorithm}{Algorithm}
\begin{algorithm}\label{algorithm:alg_1}
Carry our the following computation to determine the bare spectral strong coupling constant $\alpha_b$.
\begin{enumerate}
\item initialize $\alpha_b=1$
\item compute $d=-(\alpha_1-\alpha_n)(\Gamma_{\Lambda}(\alpha_b,\tau_1)-\Gamma_{\Lambda}(\alpha_b,\tau_n))^{-1}$
\item compute $\Delta=d-\alpha_b$
\item set $\alpha_b=d$
\item repeat steps 2, 3, and 4 until $\Delta$ vanishes
\end{enumerate}
\end{algorithm}
We have found that this algorithm converges in less than 10 iterations. The value to which it converges, i.e. the bare spectral strong coupling constant, depends on the cutoff $\Lambda$. 
The optimum value of $\Lambda$ where the running coupling agrees best with CERN $\overline{\text{MS}}$ experimental data occurs when $\Lambda$ is about 14 GeV. For this value of $\Lambda$ the Landau peak occurs at about a few hundred MeV and  the spectral bare strong coupling constant has the value 
\[ \alpha_b=0.00242962\approx\frac{1}{411}. \]

Thus it seems that we can conclude, since $\alpha_b\ll1$,  that, when analyzed using spectral regularization, QCD is perturbative at all energies, i.e. perturbation theory can be successfully used, at all energies. One can use the small finite bare coupling constant in higher order calculations in pQCD. When dimensional regularization/renormalization is used one is precluded from using the bare  coupling  constant in calculations because this quantity is infinite. 

The distance corresponding to an energy of 14 GeV is given by
\[ d=(\frac{hc}{e})(\frac{1}{14\times10^{9}eV})\approx8.86\times 10^{-17} m, \]
(where $h$ is Planck's constant, $c$ is the speed of light and $e$ is the charge of the electron) which is about of the order of the radius of a meson or baryon. 

\begin{figure} 
\centering
\includegraphics[width=14cm]{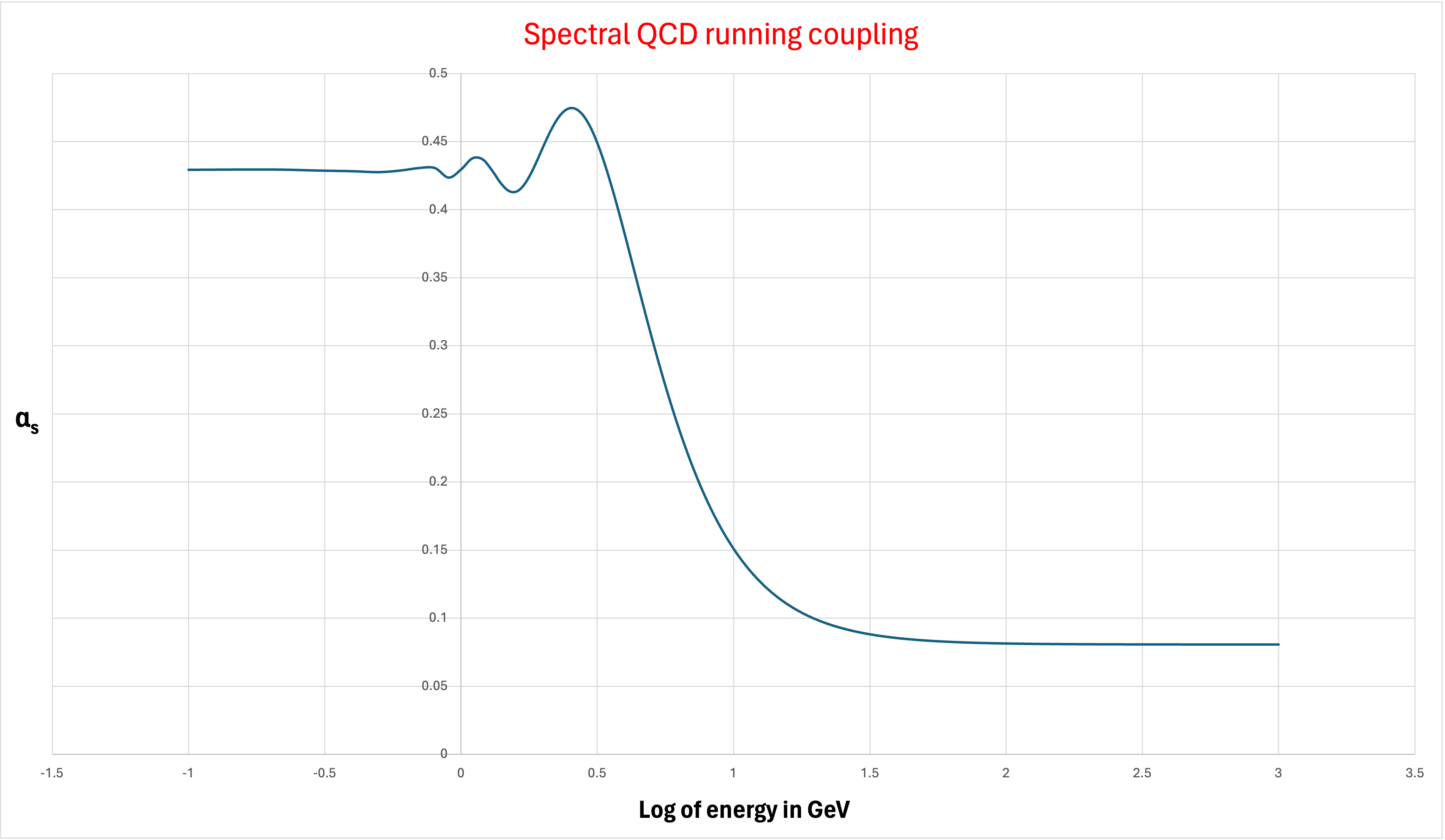}
\caption{Spectral QCD running coupling versus log of energy in GeV\label{fig:QCD_rc}}
\end{figure}

The graph of the spectral QCD running coupling $\alpha_s(Q)$ versus energy $Q$ (inverse distance) when $\Lambda=14$ GeV is shown in Fig.~\ref{fig:QCD_rc}.

It is sometimes stated that quark confinement is the result of the Landau pole (this is described as infrared slavery). However this is not correct because the Landau pole is unphysical, for example its position is renormalization scheme dependent. 

It is commonly proposed that for quark confinement to be realized a theory must have $\alpha_s(Q)\gg1$ or at least certainly $\alpha_s(Q)\geq1$ as $Q\rightarrow0$. However, confinement scenarios exist \cite{Deur2024} which do not require a large coupling e.g. the supercritical binding model discussed in Ref. \citen{Gribov} in which $\alpha_s(Q)\rightarrow0.4$ as $Q\rightarrow0$, in agreement with the limiting behavior of our $\alpha_s$.

\begin{figure} 
\centering
\includegraphics[width=14cm]{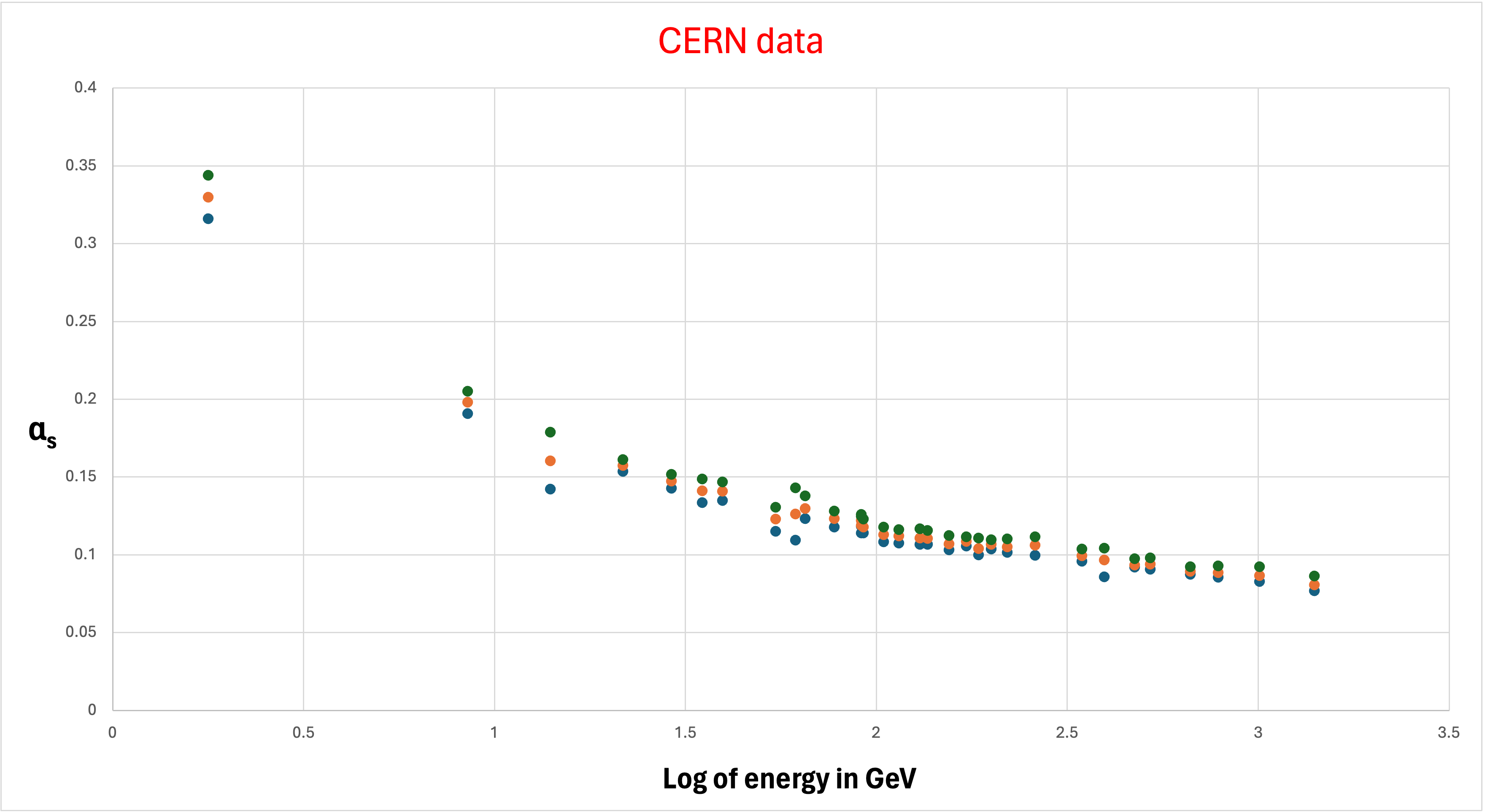}
\caption{CERN $\overline{\text{MS}}$ data. $\alpha_s$ versus log of energy in GeV.
Data courtesy of CERN\label{fig:CERN_data}}
\end{figure}

The CERN $\overline{\text{MS}}$ $\alpha_s(Q)$ versus $Q$ data is shown in Fig. \ref{fig:CERN_data}.

Our spectral running coupling has the same general shape as the $\overline{\text{MS}}$ CERN data  over the range of the CERN data with qualitative but not quantitative agreement. However the CERN data is renormalization scheme dependent, use of another renormalization scheme, other than $\overline{\text{MS}}$, in analyzing the experimental data would result in a different collection of CERN data. In fact changing from one renormalization scheme to another is associated with a transformation or mapping of the $\alpha_s(Q)$ versus $Q$ data. 

\begin{figure} 
\centering
\includegraphics[width=12cm]{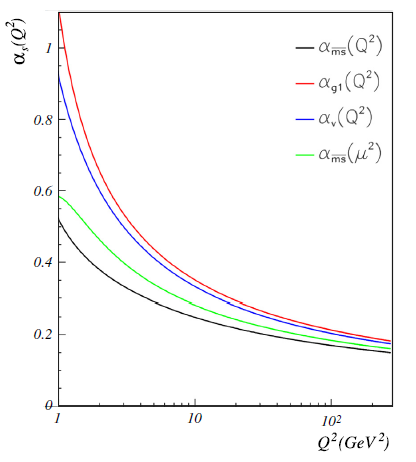}
\caption{The strong coupling $\alpha_s(Q^2)$ expressed in different renormalization schemes. {\bf Source Deur et al., 2016 \cite{Deur}, Fig. 3.4}\label{fig:Deur_Fig_3.4}}
\end{figure}

Fig.~\ref{fig:Deur_Fig_3.4} shows the QCD running coupling computed using renormalization for a number of different renormalization schemes. It can be seen that there is a range of possible QCD running coupling curves.

Spectral regularization provides a method for computing the QCD running coupling which is distinct from that associated with the $\overline{\text{MS}}$ renormalization scheme and also to all other possible renormalization schemes for the renormalization method.

We propose that if covariant spectral regularization were used to analyze the CERN raw experimental data, then the CERN data, so generated, would coincide with our theoretically computed spectral QCD running coupling curve.

\section{Freezing of $\alpha_s(\tau)$ as Energy $\tau\rightarrow0$\label{section:freezing}}

\begin{figure} 
\centering
\includegraphics[width=12cm]{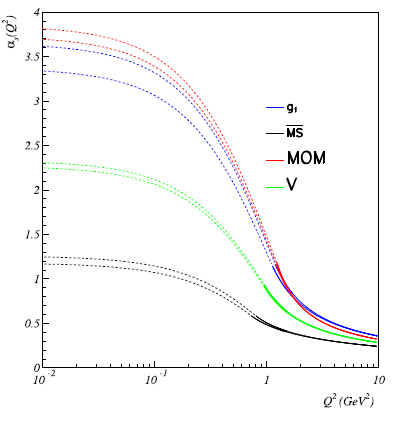}
\caption{How different renormalization schemes lead to different freezing values for $\alpha_s$. {\bf Source: Deur et al., 2016 \cite{Deur}, Fig. 5.1}   \label{fig:Deur_Fig_5.1}}
\end{figure}

At low energies the running coupling computed using the renormalization group equations in the ultraviolet in combination with standard approaches (such as $\alpha_T,\alpha_{g_1}$, or ${\chp\alpha}_{\text{PI}}$) for IR completion has the property of ``freezing" in the infrared. The freezing behavior is renormalization scheme and IR completion scheme dependent. The nature of this dependence is illustrated in Fig.~\ref{fig:Deur_Fig_5.1}. 

The spectral QCD running coupling also manifests the property of freezing at low energies as can be seen by examination of Fig.~\ref{fig:QCD_rc}. Thus, one says that spectral running coupling saturates in the IR, i.e. has an IR fixed point.

The spectral QCD running coupling does not manifest an unphysical Landau pole. It has this property in common with the spectral QED running coupling which we have shown \cite{NPB} also does not manifest a Landau pole. However the spectral QCD running coupling manifests what may be called a ``Landau peak" at an energy of a few hundred MeV, together with subsidiary subpeaks (see Fig.~\ref{fig:QCD_rc}).

\section{Conclusion}

The running coupling in QCD, while not being directly an observable (since free quarks are never or infrequently observed) is an important quantity, indicating the strength of the strong force at any given energy. It would be desirable to have a theoretical framework within which the running coupling can be computed, without {\em ad hoc} assumptions, over the whole range of energies.

Methods proposed in the literature for computing the QCD running coupling do not have the property of being rigorously formulated theories which predict the coupling in a unified way over all energies. The starting point for all the models is the renormalization group equations. The principal problem with the renormalization group equations is their prediction of an unphysical Landau pole which, for QCD, occurs at the energy of a few hundred MeV (depending on the renormalization scheme used). Therefore, proposed methods modify the renormalization group equations predictions at low energy.

For example the effective charge for the Bjorken sum rule is a modified version of the strong coupling constant at different energy scales, particularly low energy. It is determined by fitting to experimental data. Another example is the use of the AdS/CFT correspondence to extrapolate the high energy renormalization group equation strong coupling into lower energies.

The general theme of the approaches described in the literature is the interlacing of higher energy running coupling defined by the renormalization group equations with modified lower energy formulations. Our approach is not based on renormalization and hence the renormalization group equations are not applicable.

This paper focuses on spectral QCD vacuum polarization in order to compute the spectral QCD running coupling by comparing ${\mathcal M}^{\text{(tree)}}$ with ${\mathcal M}={\mathcal M}^{(\text{tree})}+{\mathcal M}^{\text{(vp)}}$ for quark $u\overline{d}\rightarrow u\overline{d}$ scattering. Using covariant spectral regularization we compute the densities associated with the quark bubble and the gluon bubble and hence the spectral vacuum polarization tensor. We then compute and display the spectral QCD running coupling.

The spectral QCD running coupling is not constructed by piecing together or matching functions defined in the perturbative and non-perturbative domains, by Pad\'{e} approximant or other means, but is defined by a single exact, well defined unified prescription over the whole CM energy range, both perturbative and non-perturbative. In fact we have shown that, when analyzed using spectral regularization, it seems that QCD is perturbative at all energies.

Based on enforcing that our running coupling curve pass through two points of the CERN $\overline{\text{MS}}$ data we compute a bare strong coupling of
\[ \alpha_b\approx\frac{1}{411}. \]
This bare coupling can be used in higher order computations in QCD. In standard pQCD one can not use the bare strong coupling in computations because it is infinite.

Our running strong coupling is a purely theoretical construct, not requiring any input data derived from experiment or lattice simulation other than the quark masses. Apart from the quark masses the only parameter which needs to be input in order to generate the running coupling function is the energy cutoff. 

The optimal energy cutoff seems, remarkably, when converted to an equivalent distance measure, to be equal to the typical hadron size.

The spectral running coupling is an analytic function which does not manifest an unphysical Landau pole but rather, what may be called a ``Landau peak" (together with subsidiary subpeaks). It has the property of freezing in the infrared.

The spectral QCD running coupling has the desirable properties of interoperability, simplicity and finiteness mentioned at the end of the article of Ref. \citen{Deur}.

In future work we will compare our results with modern experimental data and conclusions drawn from other approaches such as lattice QCD. Such work would involve processing raw experimental data or lattice QCD algorithms through the ``lens" of covariant spectral regularization (just as it is currently processed through the lens of $\overline{\text{MS}}$ regularization/renormalization) in order to compare the processed raw data or lattice QCD results with out theoretically derived $\alpha_s(Q)$ versus $Q$ curve.

We expect that if the raw QCD data were processed in this way, that is using covariant spectral regularization, then the resulting $\alpha_s(Q)$ versus $Q$ plot would agree with our theoretically derived spectral QCD running coupling  curve.

\section*{Acknowledgments}

The author thanks G\"{u}nther Dissertori and Siegfried Bethke for providing, and allowing the use of, the CERN QCD running coupling data obtained from LHC hadron collider experiments, lattice simulations, Z$^0$ pole fit, $e^{+}e^{-}$ annihilation and other sources, used to produce Figure 9.3 of the 2018  Particle Data Group Review of Particle Physics \cite{RevPartPhys}. 
The author thanks Alexandre Deur for allowing the use of two figures from his 2016 paper \cite{Deur}.
Also, the author is very grateful to two anonymous reviewers for their helpful comments and corrections.

\section*{ORCID}

\noindent John Mashford - \url{https://orcid.org/0000-0001-6100-031X}

\end{document}